\documentclass{article}
\usepackage{caption}
\usepackage{subcaption}
\usepackage{authblk}
\usepackage{graphicx}
\usepackage{amsmath}
\usepackage[a4paper, total={6in, 8in}]{geometry}

\title{Analysis of degradation in perovskite solar cells
through physics-based machine learning}

\author[1]{Kjeld O. Jensen}
\author[2]{Gemma Giliberti}
\author[2,$\dagger$]{Aldo Di Carlo}
\author[3,$\ddagger$]{Will Clarke}
\author[3]{Giles Richardson}
\author[4]{Taylor Blackwell}
\author[5]{Petra J. Cameron}
\author[4,*]{Alison B. Walker}

\affil[1]{Department of Mathematical Sciences, University of Bath, United Kingdom}
\affil[2]{CHOSE–Centre for Hybrid and Organic Solar Energy, Department of Electronic Engineering, Tor Vergata University of Rome, via del Politecnico 1, 00133 Rome, Italy}
\affil[3]{Mathematical Sciences, University of Southampton, University Rd, SO17 1BJ, United Kingdom}
\affil[4]{Department of Physics, University of Bath, United Kingdom}
\affil[5]{Department of Chemistry, University of Bath, United Kingdom}
\affil[$\dagger$]{CNR-ISM Consiglio Nazionale delle Ricerche–Istituto di Struttura della Materia, Via del Fosso del Cavaliere 100, 00133 Rome, Italy}
\affil[$\ddagger$]{School of Mathematics and Physics, University of Portsmouth, United Kingdom}

\affil[*]{Corresponding author: a.b.walker@bath.ac.uk}

\begin{document}
\maketitle
\begin{abstract}
Degradation in lead halide perovskite solar cells is analysed by inverse modelling of published measurements of characteristics of a single solar cell at ages of 0, 90, 280, and 480 minutes. We employ machine learning to deduce the distributions of material parameter values - and hence the physics - linked to the measured changes. Here, Bayesian parameter estimation is coupled with drift–diffusion simulations using the IonMonger code in combination with an optical model.  This analysis focuses on scan-rate dependent hysteresis seen in current-voltage curves and uses additional modelling to interpret photoluminescence measurements. We have accurately replicated measured changes in device performance with age through variations in model input parameters. Our key result is that degradation is influenced by correlated changes in the concentrations and diffusion coefficients of mobile ions and by interface recombination at large mobile ion concentrations. This study demonstrates the power of machine learning combined with simulations to reliably interpret experimental results, a task which is problematic if using simulation models with only manual exploration of the input parameter space. 
\end{abstract}

\section{Introduction}
Perovskite solar cells, PSCs, possess considerable advantages including greater efficiencies, low cost, low energy budget fabrication and a wide applications range. However, their lifetime (10 years for perovskite-silicon panels) is much shorter than 30-40 years for monocrystalline silicon. There is an urgent need for strategies that exploit the wide variety of perovskite compositions and device architectures in order to achieve the long-term stability required for commercial use.

 These strategies need to account for the complexity of perovskites compared to conventional semiconducting materials, such as silicon and CdTe \cite{Ceratti2026}. The best performing PSCs use hybrid organic-inorganic perovskites in which the compositions include mixed halide, triple-cation and mixed metal \cite{Blakesley2024}. Consequently there is a huge variety of chemical and physical environments in the perovskite materials, and hence in defect species, defect configuration, formation energies, diffusion constants etc.
 

There are many possible causes of degradation in PSCs, such as the intrinsic mechanisms of phase instability, left over phase impurities, phase segregation, halide segregation and ion migration, which may be influenced by external factors such as illumination \cite{Baumann2024}. While changes in material characteristics due to the increasing density of mobile ions with temperature, time, voltage and light bias have been widely studied \cite{Thiesbrummel2026}, interfaces play a crucial yet poorly understood role \cite{Wolff2019, Clarke2025,Frohna2025}. Carrier lifetimes in the perovskite layer may change and changes may also occur in the charge transport layers used to extract carriers from the perovskite. Reaching a consensus on the exact degradation pathways from the large body of measurements and linked models is therefore challenging \cite{Ceratti2026, Kirchartz2025,Kalasariya2026}.

Drift-diffusion codes that allow for ion motion, e.g. IonMonger \cite{Clarke2023jce} and DriftFusion \cite{Calado2016}, are able to model the impact of these causes of degradation on device performance and on measurements with spectroscopic techniques used to characterise devices, such as Fast-Hysteresis JV-scans and photoluminescence (PL). Measurements on a triple cation p-i-n PSC at four  device ages from 0 to 480 minutes by Thiesbrummel et al \cite{Thiesbrummel2024} combined with modelling using IonMonger \cite{Clarke2023jce}  were interpreted as showing that the dominant degradation mechanism in their devices was due to a change in a single parameter, {\it viz.}, an increase in mobile ion density, $N_0$. 
 However, more recent papers point to a more complex picture of degradation and the role of ionic defects \cite{Hart2024, Cachafeiro2025}.

A realistic model of PSC degradation typically requires more than 20 input parameters characterising the device and each of the individual layers. Given the large range of potential degradation mechanisms, many of these parameters can be expected to change as the device ages. However, a thorough exploration of parameter space incurs the curse of dimensionality \cite{Altman2018}, making a full investigation of all possible parameter combinations computationally prohibitive. It is therefore difficult to reach firm conclusions from the direct application of such models. We demonstrated a solution to this problem by using Machine Learning (ML) (specifically Bayesian Parameter Estimation (BPE)) in combination with a physics-based simulation tool (in this case IonMonger) to solve this inverse modelling problem \cite{McCallum2024}. Our approach allowed us to establish which combinations of input parameters are consistent with a given set of experimental results, e.g. a series of JV-scans of a PSC. From an analysis of hysteresis shown by the J-V-scans  we showed that mobile ion densities in the perovskite in a TiO$_2$-MAPbI$_3$-Spiro PSC were more than an order of magnitude lower than the widely cited theoretical estimate of Walsh et al \cite{Walsh2018}, instead agreeing with the estimate of Barboni and De Souza \cite{Barboni18}.

ML is an established part of atomistic calculations of materials and devices, such as those involving Molecular Dynamics (MD), Density Functional Theory (DFT) and beyond-DFT methods \cite{Arber2025, Butler2022}.  However, there are far fewer examples of physics-based ML in which PSC simulation models are combined  with techniques like Bayesian Parameter Estimation. These examples include Kober-Czerny \textit{et al} \cite{Kober-Czerny2025} on PL, Diekmann {et al} \cite{Diekmann2026} and Nabil \textit{et al} \cite{Nabil2026} using Electrical Impedance Spectroscopy, 
and our earlier work \cite{McCallum2024,McCallum2023}.

Our physics-based ML implementation reliably estimates changes in multiple model inputs and includes parameter estimates showing a high degree of correlation. Here we demonstrate its power by coupling BPE (based on Metropolis-Hastings Markov-Chain Monte Carlo) with IonMonger and RayFlare (an optical modelling package) \cite{Pearce2021}. New results are that the measurements of Thiesbrummel \textit{et al} \cite{Thiesbrummel2024} are consistent with an increase in mobile ion density $N_0$, as the device degrades. We also see a significant decrease of the mobile ion defect diffusion coefficient $D_I$, with a strong correlation between estimates of $N_0$ and $D_I$. We also investigate the effects of recombination at the interfaces between the perovskite layer and the charge transport layers (CTLs), the electron transport layer (ETL) and hole transport layer (HTL). We find evidence that the interface recombination velocity at the interface of the HTL and perovskite layer also plays a significant role in the degradation of the device characteristics. 

The next section, Results, Section \ref{sec:results},  covers the results obtained from our BPE analyses of the JV-scan results, and our extension of the analyses to open-circuit drift-diffusion simulations, which are related to photoluminescence (PL) measurements. Our analysis is given in the Discussion, Section \ref{sec:discussion}. The Conclusion, Section \ref{sec:conclusion}, summarises our easy to use method based on simulation modelling combined with ML and highlights how it can be used on experimental measurements of device characteristics to pinpoint the degradation processes occurring in the device. We describe the Physical Simulation model, the Machine Learning approach and the main input assumptions in the final Methods sections, Section \ref{sec:methods}.

\section{Methods}
\label{sec:methods}
\subsection{Physical simulation model}
\label{subsec:physsimmodel}
IonMonger is a one-dimensional drift-diffusion model for a three-layer planar PSC that accounts for the motion of electrons, holes and a single species of ion vacancy (typically the halide) within the cell. It accurately describes a variety of phenomena peculiar to PSCs including: current-voltage hysteresis \cite{Richardson2016,Calado2016,Courtier2019b}, inverted hysteresis \cite{Clarke2023JAP} and two- and three-feature impedance plots \cite{Bennett2023,Clarke2024,Jacobs2018}. Furthermore, it reproduces the results of the DriftFusion code that also allows for ion motion \cite{Calado2016}. 

The charge carrier generation profile in our simulation was calculated by an optical model instead of the IonMonger default Beer-Lambert profile \cite{Courtier2019b}. 
The RayFlare \cite{Pearce2021} package was used to implement a Transfer Matrix Method (TMM) calculation of wavelength-dependent reflection, transmission and absorption profiles in the device layers. These profiles are integrated over the standard AM1.5 100 mW/cm$^2$ solar spectrum to generate a total photon absorption profile, equivalent to the carrier generation profile in the perovskite, for which excitons initially generated are rapidly unbound. A constant multiplier, $G_{adj}$=1.184, was applied to the calculated profiles to make the simulated results compatible with the measured $J_{sc}$ values. Further details of the optical model, including the rationale for the $G_{adj}$ multiplier, are in Section S1 of the Supporting Information.

\subsection{Machine Learning}
\label{subsec:ML}
We have used the Metropolis-Hastings (MH) Markov Chain Monte Carlo (MCMC) method \cite{Gelman2013} to do Bayesian Parameter Estimation (BPE)  to establish the distribution of IonMonger input parameters that are consistent with the experimental JV-scan results. BPE applies Bayes' Rule to inversely derive a posterior distribution $p(\theta|\bf y_m)$ over a set of parameters $\theta$, associated with a measured outcome $\bf y_m$, given a prior distribution over $\theta$ and the likelihood of obtaining the measured outcome $\bf y_m$ from $\theta$, $p({\bf y_m}|\theta$). Here the measured outcome is based on the set of JV-scan results, specifically $J_{sc}$ (short-circuit current), $V_{ocr}$ (open-circuit voltage, reverse scan), $V_{ocf}$ (open-circuit voltage, forward scan), $\eta_r$ (power-conversion efficiency, PCE, reverse scan), and $\eta_f$ (PCE, forward scan), for the set of scan rates of the experimental measurements. The likelihood of obtaining a measured result $\bf y_m$ is determined from running IonMonger for a given set of parameters $\theta$ to produce a prediction ${\bf y}$, which is then compared to $\bf y_m$. For further details of the MH procedure see Section S2 of the Supporting Information.

\subsection{Device details and IonMonger inputs}
The key input to the BPE analysis is the set of experimental results from Thiesbrummel \textit{et al}  \cite{Thiesbrummel2024} on an inverted (p-i-n) cell i.e. with light coming through the HTL, using the triple-cation 1.63 eV perovskite Cs$_{0.05}$(FA$_{0.83}$MA$_{0.17}$)$_{0.95}$Pb(I$_{0.83}$Br$_{0.17}$)$_{3}$. Its architecture is ITO/PTAA/Perovskite/C$_60$/BCP/Cu. ITO is indium tin oxide, PTAA is poly[bis(4-phenyl)\\(2,4,6-trimethyl-phenyl)amine, PFN-Br is 
Poly(9,9-bis(3’-(N,N-dimethyl)-N-ethylammonium-propyl-2,7-fluorene)-alt-2,7-(9,9-dioctyl-fluorene))dibromide and BCP is bathocuproine.
JV scans for 9 scan rates at each of four different device ages (0, 90, 280 and 480 minutes). We use the values of $J_{sc}$, $V_{ocr}$, $V_{ocf}$, $\eta_r$, $\eta_f$ reported in the Supporting Information to Thiesbrummel \textit{et al} \cite{Thiesbrummel2024}. The  PFN-Br layer (less than 5 nm wide) is included to ensure close interfacial contact and achieve exceptional crystallization \cite{Wang2024}. The BCP layer (8 nm wide) is an interfacial buffer layer between the electron transport layer and the metal electrode. Neither PFN-Br nor BCP layers are explicitly considered in our model as they do not influence the optical properties due to their narrow widths. Our model considers PCBM rather than C60 as noted in SI section S1. 

For the inputs to IonMonger, unless stated otherwise, we have followed the inputs stated in \cite{Thiesbrummel2024} in their application of IonMonger.  Table \ref{tab:BPEinputs} gives the parameters included in the BPE analyses. The parameters held constant are given in Table \ref{tab:constinputs}. In the absence of refractive index data for the materials used by  \cite{Thiesbrummel2024}, we have used optical data from the device stack shown in  Figure S1 of the Supporting Information. It has similar, albeit not identical, materials \cite{Magliano2025,Giliberti2026}, and with the same layer dimensions as Reference \cite{Thiesbrummel2024}.

Reference \cite{Thiesbrummel2024} assumes carrier extraction from the Electron Transport Layer (ETL) and Hole Transport Layer (HTL) via non-ohmic contacts with cathodic and anodic work functions set to -3.95 eV and -5.45 eV respectively, giving a very high built in voltage of 1.5 V. Conventionally accepted values for $V_{bi}$ of PSCs lie below 1.25 V \cite{Hill2023}. We therefore assume ohmic contacts at the BCP/Cu extraction layer. The settings used here, as presented in Tables \ref{tab:BPEinputs} and \ref{tab:constinputs}, give $V_{bi}$ values in the range 1.0-1.5 V, depending on the value of the charge transport layer (ETL or HTL) doping densities, $d_E$ and $d_H$, when these vary in the range 10$^{21}$ to 10$^{25}$ m$^{-3}$. The values for $d_E$ and $d_H$ in the Constant column of Table \ref{tab:BPEinputs}, used in the main BPE analyses, give a $V_{bi}$ of 1.18 V.

\begin{table}
\begin{tabular}{llrrr}
\hline
{\bf Parameter} & log/lin & \multicolumn{2}{c|}{Prior range} & Constant\\
 & & low & high & \\ \hline
$N_0$, mobile ion concentration (m$^{-3}$) & log & 20 & 26 & \\
$D_I$, mobile ion diffusion constant (m$^2$/s) & log & -16 & -11 & \\
$v_{nE}$, IRV electrons  ETL (m/s) & log & -1 & 3 & \\
$v_{pE}$, IRV holes ETL (m/s) & log & -1 & 3 & \\
$v_{nH}$, IRV electrons HTL (m/s) & log & -1 & 3 & \\
$v_{pH}$, IRV holes HTL (m/s) & log & -1 & 3 & \\
$\tau_n$, electron lifetime (s) & lin & 5$\times$10$^{-8}$ & 1$\times$10$^{-6}$ & 2$\times$10$^{-7}$ \\
$\tau_p$, hole lifetime (s) & lin & 5$\times$10$^{-8}$ & 1$\times$10$^{-6}$ & 2$\times$10$^{-7}$ \\
$d_E$, doping density ETL (m$^{-3}$) & log & 21 & 25 & 3$\times$10$^{22}$ \\
$d_H$, doping density HTL (m$^{-3}$) & log & 21 & 25 & 3$\times$10$^{22}$ \\
\hline
\end{tabular}
\caption{\label{tab:BPEinputs} Input parameters included in Bayesian Parameter Estimation. The log/lin column indicates whether the BPE parameter space (and hence the prior ranges shown in the table) uses the actual value of the input ("lin") or log10 of the actual value ("log"). The Constant column shows the actual value used when a parameter is not included in the BPE, but held constant. IRV is short for Interface Recombination Velocity.}
\end{table}

\begin{table}[!]
\begin{tabular}{lr}
{\bf Parameter} & Value \\ \hline
 & \\ 
\bf{Perovskite parameters} & \\ \hline
$b$, layer thickness (m) & 500$\times$10$^{-9}$ \\
$\epsilon_p$, permittivity & 22$\times \epsilon_0$ \\
$E_c$, conduction band minimum (eV) & -3.9 \\
$E_v$, conduction band minimum (eV) & -5.52 \\
$D_n$, electron diffusion constant (m$^2$/s) & 2.59$\times$10$^{-6}$ \\
$D_p$, hole diffusion constant (m$^2$/s) & 2.59$\times$10$^{-6}$ \\
$g_c$, conduction band DoS (m$^{-3}$) & 2.2$\times$10$^{24}$ \\
$g_c$, valence band DoS (m$^{-3}$) & 2.2$\times$10$^{24}$ \\
$\beta$, bimolecular recombination rate (m$^3$s$^{-1}$) & 3$\times$10$^{-17}$ \\
$G_{adj}$, generation multiplier & 1.184 \\
 & \\ \bf{ETL parameters} & \\ \hline
$b_E$, layer thickness (m) & 30$\times$10$^{-9}$ \\
$\epsilon_E$, permittivity & 5$\times \epsilon_0$ \\
$E_{cE}$, conduction band reference energy (eV) & -3.9 \\
$g_{cE}$, conduction band effective DoS (m$^{-3}$) & 1$\times$10$^{26}$ \\
$D_E$, electron diffusion constant (m$^2$/s) & 2.59$\times$10$^{-7}$ \\
 & \\ 
\bf{HTL parameters} & \\ \hline
$b_H$, layer thickness (m) & 10$\times$10$^{-9}$ \\
$\epsilon_H$, permittivity & 3.5$\times \epsilon_0$ \\
$E_{vH}$, valence band reference energy (eV) & -5.5 \\
$g_{vH}$, valence band effective DoS (m$^{-3}$) & 1$\times$10$^{26}$ \\
$D_H$, hole diffusion constant (m$^2$/s) & 2.59$\times$10$^{-9}$ \\
\hline
\end{tabular}
\caption{\label{tab:constinputs} Input parameters held constant in the IonMonger simulations. Except for the cases discussed in the text, these are based on the values used in Ref. \cite{Thiesbrummel2024}. $\epsilon_0$ is the vacuum permittivity. $G_{adj}$ is explained in Section \ref{subsec:physsimmodel}.}
\end{table}

\begin{table}
\caption{Parameter estimates, derived from the means of BPE posterior distributions for the four device ages. Figures in italics are means of distributions without a clearly defined peak and hence strongly influenced by the choice of priors.}
\centering
\begin{tabular}{l c c c c}
\hline
Input & Age 0 min & Age 90 min & Age 280 min& Age 480 min \\ \hline    
$N_0$ (m$^{-3}$) & 3.76$\times$10$^{21}$ & 1.61$\times$10$^{23}$ & 1.79$\times$10$^{23}$ & 8.70$\times$10$^{23}$ \\
$D_I$ (m$^2$s$^{-1}$) &  2.68$\times$10$^{-13}$ & 1.85$\times$10$^{-14}$ & 2.45$\times$10$^{-14}$ & 7.01$\times$10$^{-15}$ \\
$v_{nE}$ (ms$^{-1}$) & \it{10.65} & \it{17.54} & \it{3.02} & \it{5.47} \\
$v_{pE}$ (ms$^{-1}$) & \it{10.7} & \it{14.2} &  \it{8.0} & \it{11.3}  \\
$v_{nH}$  (ms$^{-1}$)& 2.5 & 5.1 & 8.0 & 16.5 \\
$v_{pH}$  (ms$^{-1}$)& 8.8 & 3.0 & 2.8 & 2.4 \\
\hline
\end{tabular}
\label{tab:BPEmeanssd}
\end{table}

\section{Results}
\label{sec:results}

This section focuses on the results of the BPE analyses. In addition, we report on a series of individual IonMonger runs that explore the sensitivity of the JV-scan results to different input parameters and on runs at open-circuit related to PL measurements. More detailed descriptions, results and discussion can be found in the Supporting Information (SI).

The BPE analysis proceeds in a sequence of steps, initially estimating the distributions for a larger number of model input parameters, then reducing the number of these parameters by choosing the parameters where the BPE posterior distributions  (parameter probability distributions updated by the ML so device characteristics agree as far as possible with experiment) show that their values have a discernible effect on the JV-scan results.

The initial set of BPE inputs comprises the following model parameters: mobile ion density $N_0$, mobile ions diffusion constant $D_I$, recombination velocities of carriers (suffix $n$ for electrons, suffix $p$ for holes) at the perovskite-CTL interfaces (suffix $E$ for ETL, suffix $H$ for HTL) $v_{nE}$, $v_{pE}$, $v_{nH}$, $v_{pH}$, carrier lifetimes for electrons and holes, respectively, in the perovskite, $\tau_n$, $\tau_p$, and doping densities in the ETL and HTL, respectively, $d_E$, $d_H$.
This set of model parameters was chosen to test hypotheses about which material and device properties change during degradation: mobile ion properties, interface recombination, carrier lifetimes and charge transport layer properties. The recent review by Thiesbrummel \textit{et al}  \cite{Thiesbrummel2026} quotes these device parameters as the main candidates for explaining performance degradation in aging PSCs.

Our BPE analyses which included the parameters $\tau_n$, $\tau_p$, $d_E$, and $d_H$ found that the posterior distributions for these parameters were close to uniform across the prior ranges, did not show convergence towards a narrower range of values, nor show any substantial correlation with the estimates of other parameters, nor obvious signs of trends associated with device aging. Moreover, the results of these BPE analyses do not point to any different conclusions from the conclusions of the main BPE analyses presented below. Figures S3 and S4 in the SI show BPE analyses conducted with this wider set of parameters. These figures show that the inclusion of these extra parameters does not make a significant difference to the JV-scan results. For the main set of BPE analyses the parameters were set to constant values; for both $\tau_n$ and $\tau_p$ we used the values employed in Thiesbrummel \textit{et al}, \textit{i.e.}$2 \times 10^{-7}$s, whilst for both $d_E$ and $d_H$ we used a value of $3 \times 10^{22}$m$^{-3}$. 

\begin{figure}
 \centering
 \includegraphics[width=1.0\textwidth]{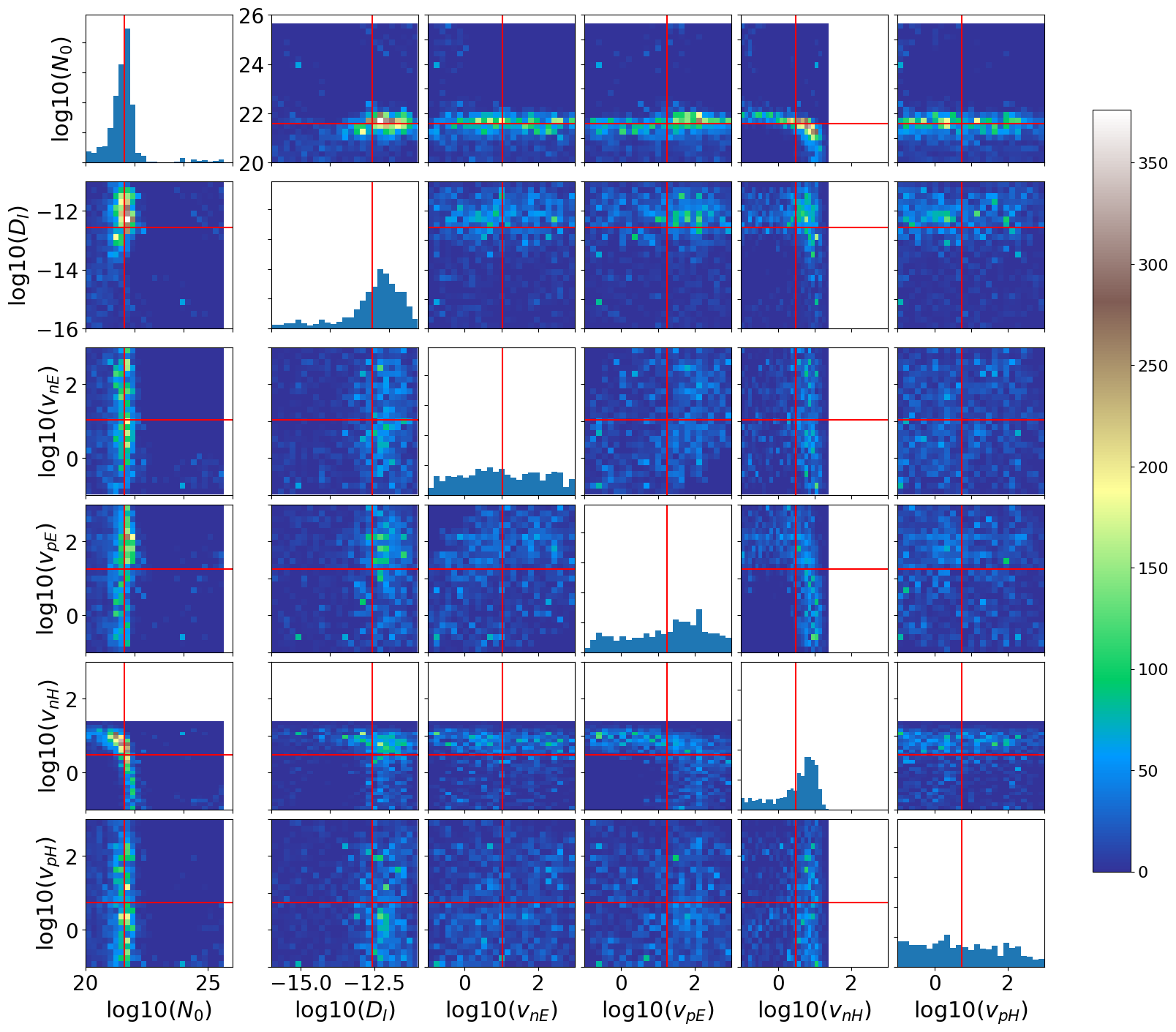}
    \caption{Posterior distributions (counts of samples from the MCMC chains) from the BPE analysis of the age-0-min device, showing single parameter distributions along the diagonal (all with the same y-scale) and joint two-parameter distribution as heat maps off-diagonal (with each joint distribution appearing twice, on either side of the diagonal). The common colour scale used for the heat maps are shown in the colour bar. The red lines show the means of the (single-variable) distributions. The quantities shown are log$_{10}$ of the input parameters, $N_0$, $D_I$, $v_{nE}$, $v_{pE}$, $v_{nH}$, $v_{pH}$, in SI units. The limits of the axes for each parameter correspond to the prior-ranges of that parameter. Thus white space shows regions where there are no samples.  }
    \label{fig:age0allpost}
\end{figure}

\begin{figure}
    \centering
    \includegraphics[width=1.0\textwidth]{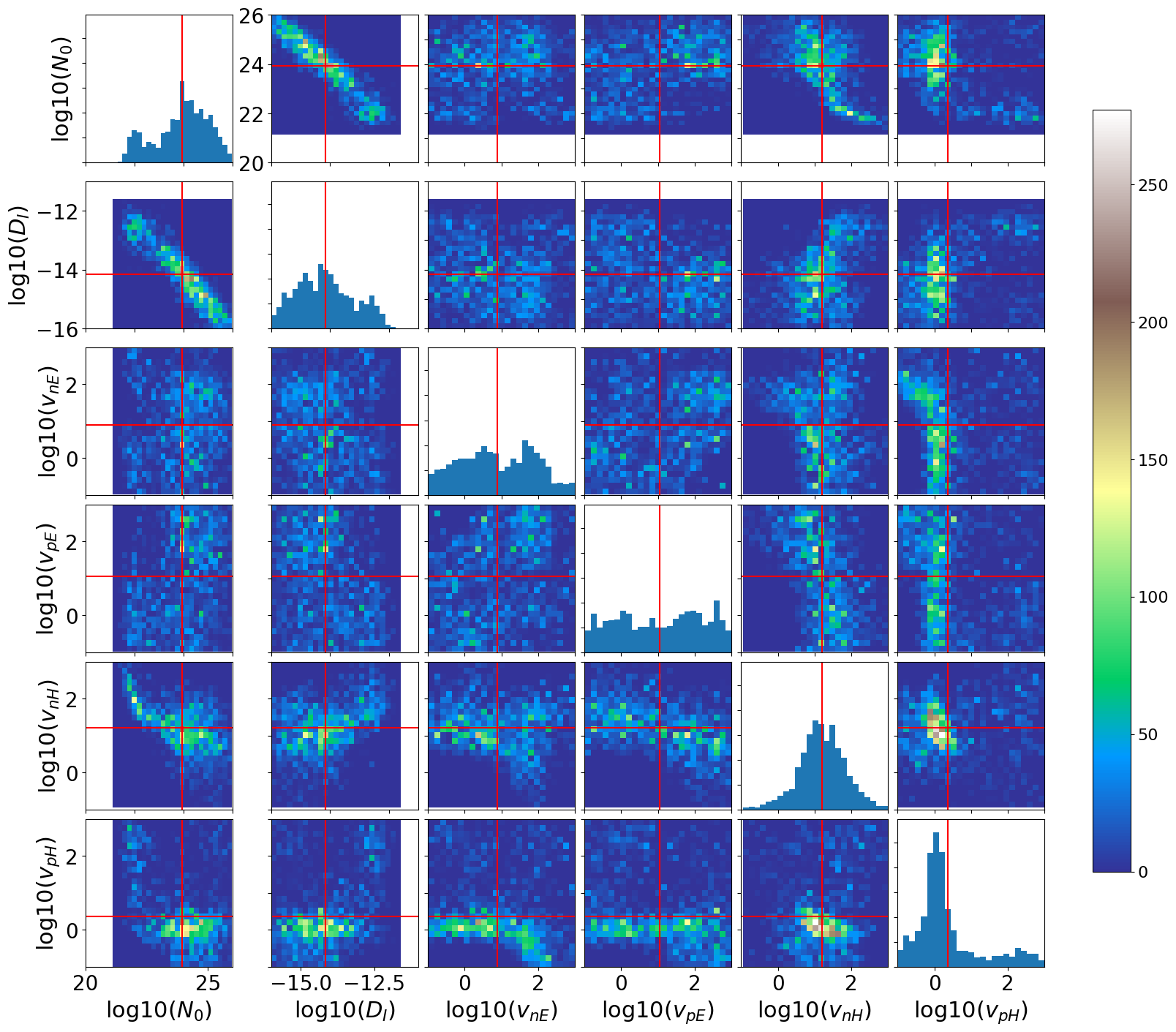}
    \caption{Posterior distributions from the BPE analysis of the age-480-min device. See caption to Figure \ref{fig:age0allpost} for description. }
    \label{fig:age480allpost}
\end{figure}

\begin{figure}
    \centering
    \includegraphics[width=1.0\textwidth]{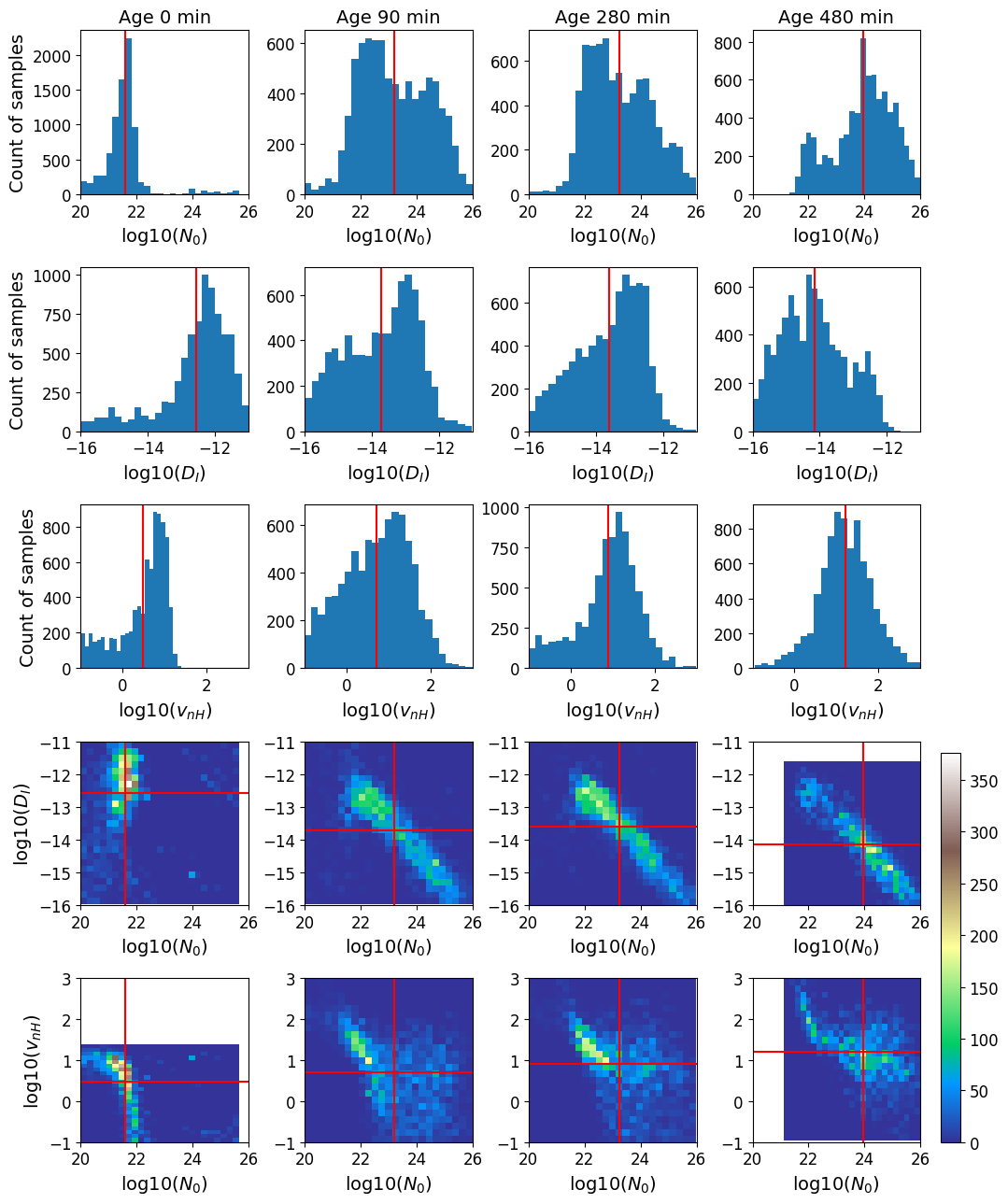}
    \caption{Posterior distributions from BPE for the four ages for selected input parameters, showing single parameter distributions for $N_0$, $D_I$ and $v_{nH}$ along the top three rows and joint two-parameter distribution for $N_0$-$D_I$ and $N_0$-$v_{nH}$ as heat maps in the bottom two rows. See caption to Figure \ref{fig:age0allpost} for additional description.}
    \label{fig:allagepost}
\end{figure}

The core BPE analysis was thus focused on the six input parameters: $N_0$, $D_I$, $v_{nE}$, $v_{pE}$, $v_{nH}$, and $v_{pH}$. The posterior distributions for the 0 min and 480 min aged devices are shown in Figures \ref{fig:age0allpost} and \ref{fig:age480allpost}, respectively. These figures show both the marginal distributions of individual parameters and the pairwise joint distributions, presented as heat maps. Figure \ref{fig:allagepost} shows selected marginal and pairwise joint distributions for all four aging times. Table \ref {tab:BPEmeanssd} presents the parameter estimates, derived by means of the BPE posterior distributions, translating the log10 values used in the BPE analysis to the actual input values. The statistical errors in estimating the means are small due to large effective sample sizes (see Section S2 of the Supporting Information). However, this analysis does not account for any epistemic errors, i.e. those arising from the model assumptions.

Figure \ref{fig:age0allpost} shows that the analysis of the JV-scan results for the pristine device produces well-defined peaks in the posterior distributions for $N_0$, $D_I$ and $v_{nH}$. Table \ref{tab:BPEmeanssd} shows that the estimates of $N_0$ increase by a factor of about 200, from $3.76 \times 10^{21}$ to $8.7 \times 10^{23}$ m$^{-3}$, as the device ages from 0 to 480 minutes. Moreover, the estimates of $N_0$ and $D_I$ are strongly inversely correlated, and $D_I$ decreases by a factor of 40 over the same age range. 

Although in Figure \ref{fig:age480allpost} the posterior distributions of $N_0$ and $D_I$ at ages greater than 0 min are individually quite broad, there is a sharp ridge in the pair-wise joint $N_0$-$D_I$ distribution, as seen in top-left panels of this Figure for the age-480-min device and in the $N_0$, $D_I$ distributions. The position of this ridge in the $N_0$-$D_I$ parameter space does not change with age, but the weight of the distribution shifts along the ridge from low-$N_0$/high-$D_I$ to high-$N_0$/low-$D_I$. 

\begin{figure}
    \centering
    \includegraphics[width=0.8\linewidth]{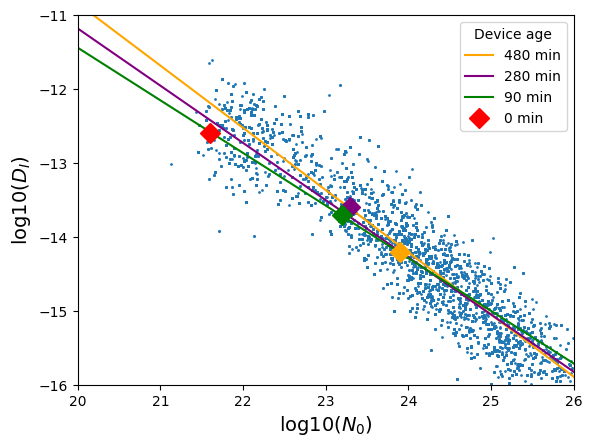}
    \caption{Scatter plot of log10($D_I$) vs log10($N_0$) from BPE samples for the age-480-min device (blue), linear fit to these points (orange) and the linear fits to similar scatter plots for the ages 90 min and 280 min devices (green and purple lines, respectively). The diamonds show the points determined by the means of N0 and DI for each of the ages, including the red diamond for the age-0-min device.}
    \label{fig:scatterslopes}
\end{figure}

Figure \ref{fig:scatterslopes} shows that when lines are fitted to the 90, 280 and 480 min results, they lie close to each other with slopes of $\approx$ -0.8. 
This result implies that the JV-scan results are much more sensitive to the ion {\it conductivity} ($\propto N_0\times D_I$) than to $N_0$ and $D_I$ individually and that the ion conductivity varies much less with age than $N_0$ and $D_I$ individually. Our results demonstrate that for the model to remain consistent with the experimental results, the $N_0$ and $D_I$ values have to be near the ridge seen in the heat maps. Any combination of $N_0$ and $D_I$ values that is {\it not} in the vicinity of this ridge would not be consistent with experiment, including $N_0$-$D_I$ combinations where only $N_0$ changes. 

This observation can be at least partially explained by the findings of \cite{Richardson2016,Courtier2019a,Courtier2019b}, who show in their derivations of several versions of the Surface Polarisation Model that for most PSCs under standard operating conditions the ionic problem is almost completely decoupled from the charge-carrier problem. Moreover, the temporal response of the cell is largely determined by the solution of this ionic problem. Matching the characteristic timescale for changes in cell behaviour to that observed experimentally thus leads to the strong correlation between the two key ionic parameters, $D_I$ and $N_0$. Notably, no such correlation between $N_0$ and $D_I$ is observed in the undegraded device. This can be attributed to the fact that $N_0$ is sufficiently low (in the undegraded device) that the Debye length in the perovskite is comparable to the layer thickness, invalidating the assumptions that underlie the Surface Polarisation Model and resulting in a device with very small hysteresis. 

The estimates for the interface recombination velocities at the HTL-perovskite interface change with age, as seen in the $v_{nH}$ and $v_{pH}$ panels of Figures \ref{fig:age0allpost}-\ref{fig:allagepost}: $v_{nH}$ (interface recombination velocity for electrons) increases and is correlated with $N_0$, whereas $v_{pH}$ (interface recombination velocity for holes) goes from a broad posterior distribution for the age-0 device, indicating that the value is only weakly related to the JV-scan results, to having a prominent peak near 1 ms$^{-1}$. This result shows that $v_{pH}$ becomes more important as mobile ion screening increases, due to higher hole concentrations at the perovskite-HTL interface. In contrast, as seen in the $v_{nE}$ and $v_{pE}$ panels of Figures \ref{fig:age0allpost}-\ref{fig:age480allpost}, interface recombination velocities at the perovskite-ETL interface always show very broad, in some cases essentially flat, distributions, indicating that these parameters do not have a substantial impact on the JV-scan curves.

\begin{figure}
    \centering
    \includegraphics[width=1\linewidth]{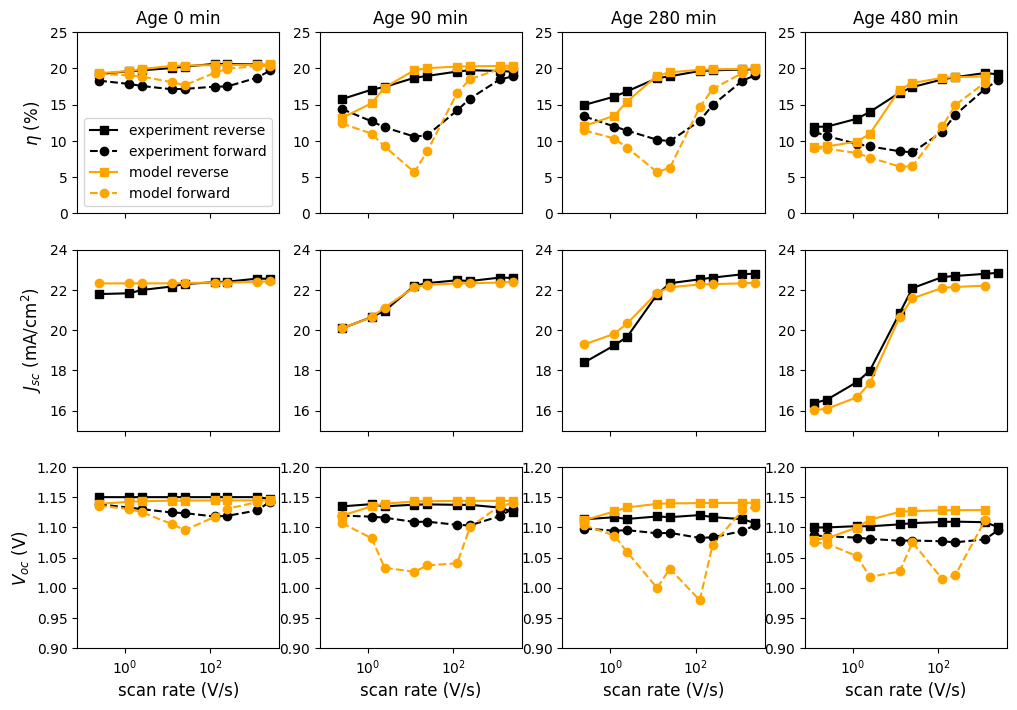}
    \caption{IonMonger JV-scan results (orange) with inputs set to the means from the BPE (Table \ref{tab:BPEmeanssd}), compared to the experimental results from reference \cite{Thiesbrummel2024} (black). Solid lines denote the reverse scans, dashed lines the forward scans. The four columns correspond to the four device ages. The top row show the power conversion efficiency ($\eta$), the second row the short-circuit current ($J_{sc}$), the third row the open-circuit voltage ($V_{oc}$). . }
    \label{fig:JVmodelvsexp}
\end{figure}

\begin{figure}
    \centering
     \includegraphics[width=1\linewidth]{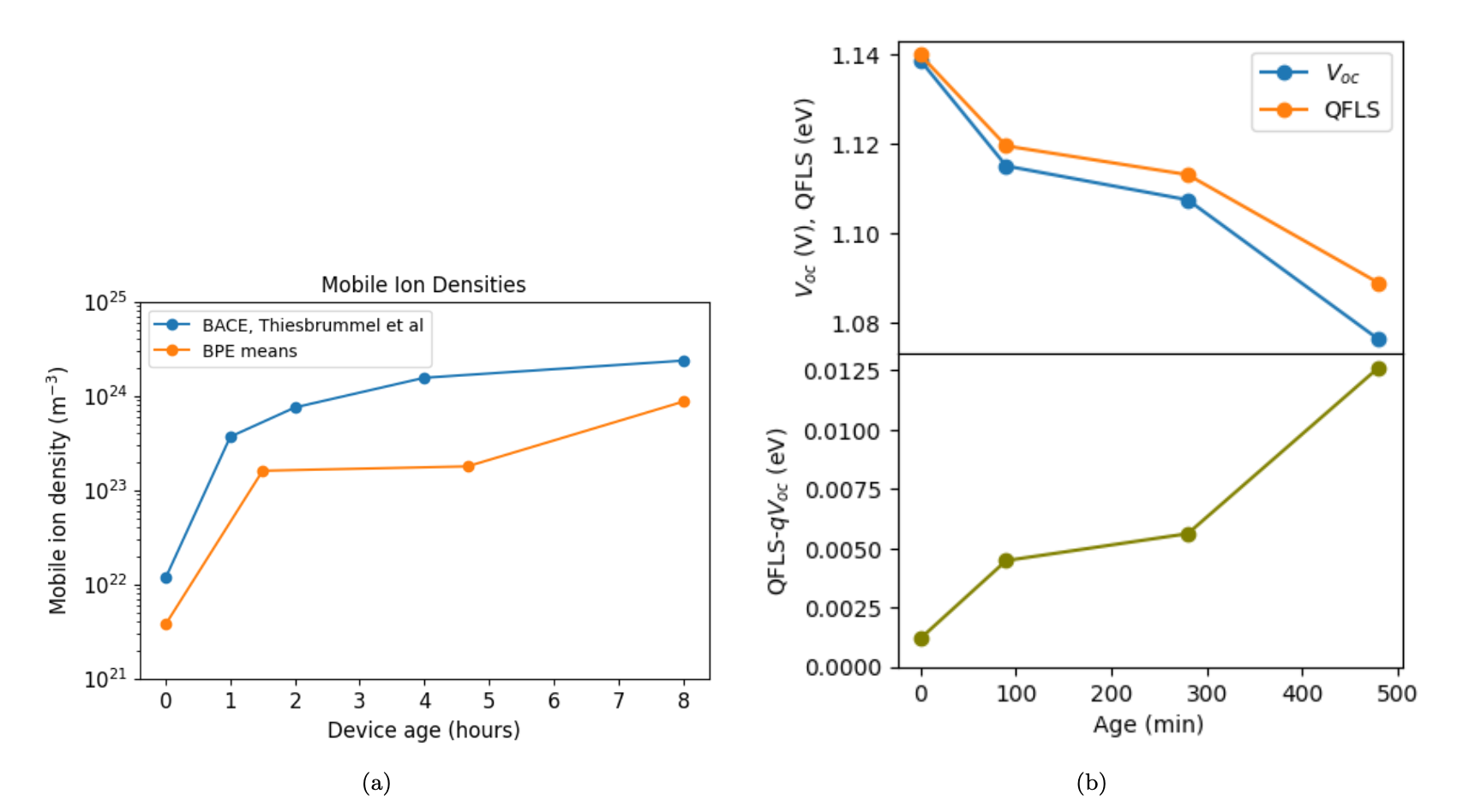}
    \caption{Panel (a): Comparison of Mobile Ion Densities: BPE means from the current analysis (orange line and points) and the densities derived by Thiesbrummel {\it et al} from Bias Assisted Current Extraction (BACE) measurements \cite[including the value at age 0 from their Supporting Information]{Thiesbrummel2024} (blue line and points). Panel (b): Quasi-Fermi-Level Splitting (QFLS) (orange line and points) and open-circuit voltage ($V_{oc}$) (blue line and points) from different open-circuit runs of IonMonger. Top half shows the individual values of QFLS and $V_{oc}$, bottom half their difference (green line and points).}
    \label{fig:N0_BACE_BPE}
\end{figure}

Running IonMonger with parameters determined by the BPE  provides good agreement with the experimental JV-scan results and in particular replicates the observed trends as the device ages. Figure \ref{fig:JVmodelvsexp} shows IonMonger results for all 4 ages compared to the experimental results. As seen in Figure \ref{fig:N0_BACE_BPE}(a), the age dependence of the mean mobile ion densities ($N_0$) agrees well with that reported by Thiesbrummel {\it et al} \cite{Thiesbrummel2024}, derived from Bias-Assisted Current Extraction (BACE) measurements. Reference \cite{Thiesbrummel2024} expresses reservations about the quantitative accuracy of their BACE $N_0$ values, since the sensitivity of BACE to variations in $N_0$ above about 2$\times$10$^{22}$m$^{-3}$ is expected to be limited \cite{Diekmann2023}. Nevertheless, the values they quote are close to what we derive from the JV-scans. From the BACE measurements, Reference \cite{Thiesbrummel2024} reports a single value of the ion diffusion constant of $7 \times 10^{-14}$m$^2$s$^{-1}$, which is within the range of values reported here, but they did not consider any age dependence of $D_I$. 

To explore the influence of individual inputs on the JV-scan results, we made a series of IonMonger runs separate from the BPE analysis. The results are presented in Section S4 of the Supporting Information alongside a detailed discussion of each step of this analysis. Our key observations are as follows.

Increasing $N_0$ as the device ages leads to more hysteresis and a drop in $J_{sc}$ at low scan rates. If this increase were not accompanied by a decrease in $D_I$, the scan rate where the greatest change in $J_{sc}$ occurs would increase as $N_0$ increases. This result is not seen in the experimental results as the device degrades and is the origin of the $N_0$-$D_I$ "ridge" seen in the BPE results. The drop in $J_{sc}$ as scan rates vary from fast to slow was also seen in the results of Cave {\it et al} \cite{Cave2020}, who showed that the scan rate at which hysteresis peaks is directly related to $D_I$.

An increase in $N_0$ correlated with a decrease in $D_I$, without changing any other parameters, can produce results that are fairly close to the experimental results. 
However, a better fit with experiments comes from also changing interface velocities, in particular increasing $v_{nH}$ in addition to $N_0$ and $D_I$. The results show that the hysteresis and the drop in $J_{sc}$ at low scan rates is highly sensitive to changes in $v_{nH}$, but only when $N_0$ values are high. This effect is readily understood because the field screening from the mobile ions reduces the internal field. Therefore the carrier densities will be more uniformly distributed across in the perovskite layer, leading to higher electron concentrations at the HTL-perovskite interface, increasing the impact of changes to the recombination parameters at that interface. Accordingly, increases in interface recombination cannot, on their own, explain the changes seen in the JV-scans as the device degrades, as observed in reference \cite{Thiesbrummel2024}. 

Finally, we performed a series of IonMonger simulations at open-circuit voltage, using as inputs the parameter values obtained from the BPE analyses for the four device ages described in Section S3 of the Supporting Information. These simulations allow us to calculate the mean of the Quasi-Fermi-Level Splitting (QFLS) across the device and compare its value to $qV_{oc}$ ($q$ is the elementary charge), which can be related to the PL results reported in \cite{Thiesbrummel2024}. As shown in Figure \ref{fig:N0_BACE_BPE}(b), the simulations confirm that for the pristine device, QFLS and $qV_{oc}$ have nearly identical values, but a gap between QFLS and $qV_{oc}$ opens up as the device degrades. The results indicate that the increase in $N_0$ {\it on its own} does not lead to a substantial QFLS-$qV_{oc}$ gap; the growing gap is only seen when the changes in interface recombination velocities, estimated from the BPE, are included in the simulations. We thus are able to replicate the key features of the reported PL results. However, there are quantitative differences: (i) the simulations estimate a QFLS-$qV_{oc}$ gap of 0.012 eV for the Age-480-min degraded device, whereas the results in Ref. \cite{Thiesbrummel2024} show values greater than 0.1 eV, (ii) in the simulations {\it both} QFLS and $qV_{oc}$ go down with device age, whereas reference \cite{Thiesbrummel2024} reports QFLS {\it increasing} with device age. We find that it is not possible to replicate both of those features simultaneously in the simulations using any combination of $N_0$, $D_I$ and the four interface recombination velocities.  

\section{Discussion}
\label{sec:discussion}
Our results have demonstrated the power of combining simulation models, such as IonMonger, with Bayesian Parameter Estimation, specifically in the study of degradation in Perovskite Solar Cells. Our analysis of the Fast Hysteresis JV-scan results of Thiesbrummel {\it et al} \cite{Thiesbrummel2024} clearly show that the mobile ion density ($N_0$) is increasing as the device degrades. However, there is strong evidence, not reported in Reference \cite{Thiesbrummel2024}, that in addition the mobile ion diffusion constant ($D_I$) decreases with device age. Furthermore, our results also show evidence of changes n the interface recombination velocities at the perovskite-CTL interfaces. In particular, the JV-scan results are strongly sensitive to changes in the interface recombination velocities in the presence of large mobile ion densities.
Our results agree with the main conclusion of reference \cite{Thiesbrummel2024}, i.e. there is an increase in $N_0$ as the device degrades. It is well established that mobile ions influence cell outputs in part through field screening \cite{Thiesbrummel2026}, e.g. current-voltage hysteresis seen at widely used scan rates \cite{Cave2020, McCallum2024} and Warby {\it et al} \cite{Warby2023} come to a similar conclusion.  However, our analysis reveals a more complex picture of degradation. Aging is associated not only with changes in the iodide vacancy density $N_0$ and diffusivity $D_I$, but also with changes at the perovskite–CTL interfaces. These effects are coupled: our estimated changes in $N_0$ and $D_I$ are correlated and the influence of the interfaces becomes stronger for higher values of $N_0$. As illustrated in Figure \ref{fig:scatterslopes}, the coupling between $N_0$ and $D_I$ is reflected in the fact that the product $N_0 D_I$ varies much less than $N_0$ and $D_I$ do individually, as the device degrades. Indeed, it is remarkable that there appears to be a power-law relation between $N_0$ and $D_I$ (as illustrated in Figure \ref{fig:scatterslopes}) as the device degrades. 


We therefore argue that labelling any difference between PCE at very high scan rates ("ion freeze") and at very low scan rates ("steady state") purely as "ionic loss", as proposed in reference \cite{Thiesbrummel2024}, does not capture the full complexity of what is occurring in any particular device. Cachafeiro {\it et al} \cite{Cachafeiro2025} observe that the problem with this "ionic loss" metric (acknowledged by \cite{Thiesbrummel2024}) is that at fast scan rates, mobile ions can influence the PCE even when they are not moving during the scan, depending on their concentration profile across the device and at the interfaces at the beginning of the scan. Consequently, as noted by Cachafeiro {\it et al} \cite{Cachafeiro2025}, the PCE difference between a fast and a slow JV scan is a potentially flawed diagnostic for the detection of pure mobile ion effects compared to other sources of degradation. 

Other studies have argued that other processes contribute to the degradation in performance. In earlier work \cite{Clarke2025} we have shown that experimental Electrical Impedance Spectroscopy (EIS) measurements on aging PSCs can be reproduced by introducing an interface recombination rate that increases with time. In this case an increase in mobile ion density did not give results that closely resembled experiment. Rombach {\it et al} \cite{Rombach2024} attribute degradation in Pb-Sn PSCs not just to increased impact of the redistribution of mobile ions during device operation but also to further losses from increases in non-radiative recombination and background hole density (the latter being a feature in Pb-Sn PSCs, but less so in Pb-only PSCs). Frohna {\it et al} \cite{Frohna2025} showed that engineering stable interfaces is critical to achieving robust devices in a study of double-cation PSCs.

For mixed-halide perovskites, the study by Kalasariya {\it et al} \cite{Kalasariya2026} has also shown that halide segregation \cite{Knight2020}, where an initial random mixture of halide atoms segregates into regions dominated by one or other of the halide constituents, can occur on similar time scales to those studies, i.e. $<$1000 minutes for 83:17 I:Br perovskite devices, and that this segregation can have a substantial impact on experimental results, in particular PL. We discuss this source and other potential sources of the discrepancy between our modelled results for the PL QFLS and those presented in Reference \cite{Thiesbrummel2024} in Section S3 of the Supporting Information and Figures S5, S6.


The range of values from 7.1$\times$10$^{-15}$ to 2.7$\times$10$^{-13}$ m$^2$s$^{-1}$ for the ion diffusion constant, $D_I$, estimated from our BPE analyses, is consistent with the range of values quoted in the recent review by Thiesbrummel {\it et al} \cite{Thiesbrummel2026}, drawing from a wide range of different sources. However, there is no discussion in \cite{Thiesbrummel2026} related to changes in $D_I$ on degradation or any pointers to how an increase in mobile ion density may lead to a lower effective diffusion constant for these mobile ions. Both Barboni and de Souza \cite{Barboni18,DeSouza19} and Moia and Mayer \cite{Moia2026} have emphasised that ionic conductivity in perovskites is highly dependent on illumination, although this dependence does not in itself explain how continued illumination can lead to higher ion concentration over time, nor how the mean ion diffusion constant may change. Other authors, such as \cite{Diethelm2025, Leupold2021}, have commented on the fact that ionic conductivity is key to understanding PSC characteristics.

Several studies, e.g. \cite{Yantara2024}, discuss the possibility that there may be a large distribution of diffusion constants for any particular ionic species as well as multiple mobile ion species.  The results of Kumar {\it et al} \cite{Kumar2026} from capacitance-frequency spectroscopy for PSCs under thermal stress at 85$^\circ$ C, and those of Reichert {\it et al} \cite{Reichert2020} from temperature-dependent deep-level transient spectroscopy (DLTS) measurements both point to this conclusion. Therefore, the population of mobile defects in the degraded device, seen in the experimental measurements reported in \cite{Thiesbrummel2024}, may be different from those in the pristine device, in particular given the complex composition of the perovskite (triple cation, mixed halide). With these observations in mind we note that what is modelled by IonMonger as a single ionic species may be the aggregate result of multiple types of defects, and the values estimated by BPE would thus be averages or aggregations. In principle, drift-diffusion modelling can deal with any number of different mobile charged defect species. However, given the experimental evidence currently available, it is difficult to define all the necessary inputs for such a model.

\section{Conclusions}
\label{sec:conclusion} 
In this paper we have demonstrated how Bayesian Parameter Estimation using Markov Chain Monte Carlo can be used in the interpretation of current-voltage scan measurements on perovskite solar cells (PSCs), with a particular focus on analysing changes in devices as they age and experience performance degradation. This study has shown the advantage of using BPE to explore the input parameters space for physical simulations such as the drift-diffusion model IonMonger, for multilayer devices, where models inevitably have a large number of inputs. Without BPE or an alternative ML technique it will always be hard to reliably interpret experimental measurements on PSCs or similarly complex systems using simulation models.

This analysis has allowed us to construct a complex picture of degradation with changes to multiple quantities: mobile ion densities, mobile ion diffusion constants, and carrier recombination parameters. This suggests a degradation process that involves multiple defect species or defect species in different configurations within the perovskite, as well as changes to the charge transport layer-perovskite interfaces. New results to our knowledge are firstly that aging is associated not only with changes in the iodide vacancy density $N_0$ and diffusivity $D_I$, but also with changes in recombination velocities at the perovskite–transport layer interfaces; and secondly that there appears to be a power-law relation between $N_0$ and $D_I$ (as illustrated in Figure \ref{fig:scatterslopes}) as the device degrades 

Our application of the combination of a simulation with Bayesian Parameter Estimation is a point on a trajectory towards a physics-based machine learning model that runs fast enough to update at the same rate as changes occur, a model that can be operated effectively as a digital twin \cite{Wright2020} of, for example, a perovskite solar cell. This would be particularly attractive for PSCs, as they experience greater variation in performance over time than, {\it e.g.} silicon solar cells. Furthermore, our approach is readily adapted to outdoor or operational monitoring of PSC degradation, including possible implications for long-term energy-yield prediction.

\subsection*{Conflict of Interest}
The authors declare that they have no conflict of interest.

\subsection*{Acknowledgements}
We are grateful to Nicola Courtier for useful discussions on how to integrate IonMonger with optical models. 
Giliberti, Di Carlo thank the European Union Horizon Europe Energy program, project 101122288—SolMates.


\bibliographystyle{unsrt}
\bibliography{perovskites.bib}
\appendix
\section{Supplementary Information}
\renewcommand\thefigure{\thesection.\arabic{figure}}
\setcounter{figure}{0}

\subsection{Calculation of charge generation profiles by optical modelling}
\label{sec:SIopt}

\begin{figure}
\centering
\begin{subfigure}{0.5\textwidth}
    \centering
    \includegraphics[width=.60\linewidth]{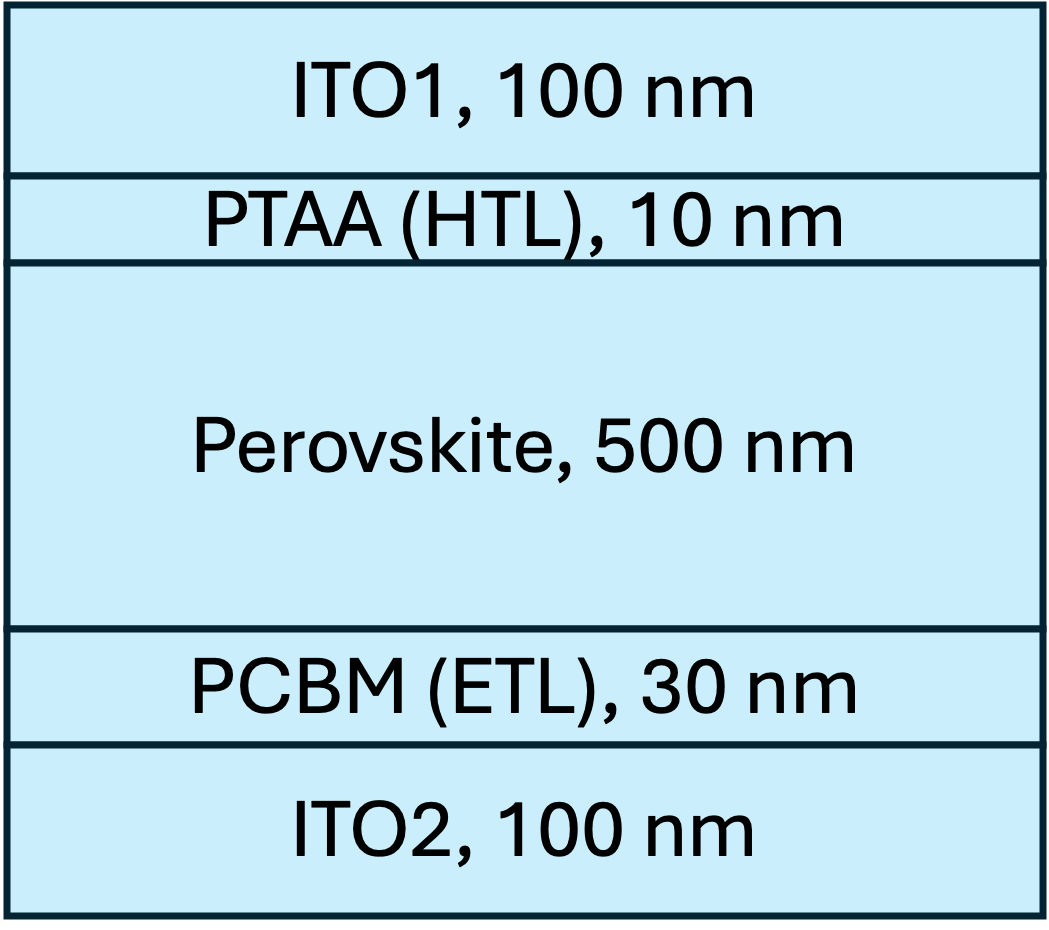}  
    \caption{}
    \label{SUBFIGURE LABEL 1}
\end{subfigure}
\begin{subfigure}{0.45\textwidth}
    \centering
    \includegraphics[width=.95\linewidth]{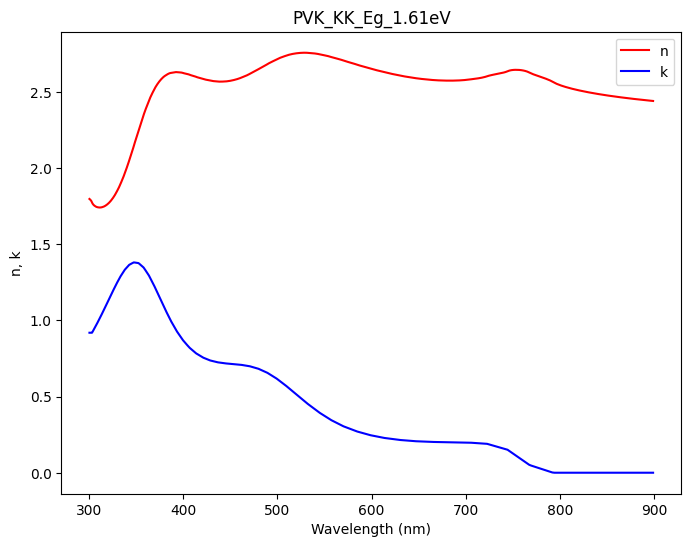}  
    \caption{}
    \label{SUBFIGURE LABEL 2}
\end{subfigure}
\begin{subfigure}{0.45\textwidth}
    \centering
    \includegraphics[width=.95\linewidth]{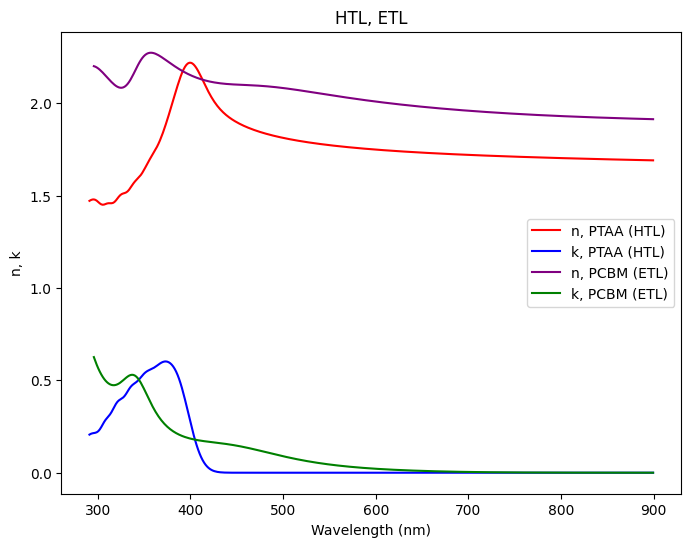}  
    \caption{}
    \label{SUBFIGURE LABEL 3}
\end{subfigure}
\begin{subfigure}{0.45\textwidth}
    \centering
    \includegraphics[width=.95\linewidth]{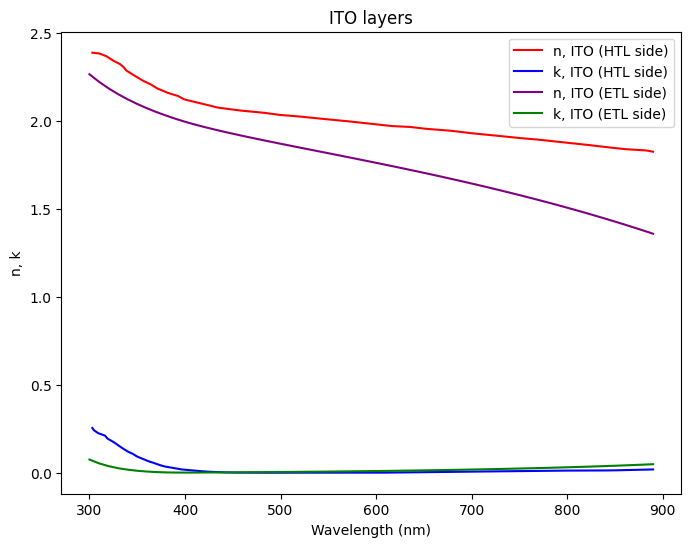}  
    \caption{}
    \label{SUBFIGURE LABEL 4}
\end{subfigure}
\caption{Device stack for the TMM calculation and wavelength variation of the refractive indices, n and k, for each of the layers \cite{Magliano2025, Giliberti2026}. Illumination is on the HTL side. }
\label{fig:stack_nk2}
\end{figure}

Figure \ref{fig:stack_nk2} shows the device stack and the refractive indices of each of the layers. 
The charge carrier generation profile in our simulations was calculated by an optical model, instead of using the IonMonger default Beer-Lambert profile \cite{Courtier2018}. RayFlare software \cite{Pearce2021} was used to implement a Transfer Matrix Method (TMM) calculation of wavelength-dependent reflection, transmission and absorption profiles in the different layers of the device. These profiles were integrated over the AM1.5 100 mW/cm$^2$ solar spectrum to generate a total photon absorption profile. Each absorbed photon generates an electron-hole pair which unbinds, producing free electrons and holes.

The thicknesses of the perovskite and charge-transport layers in our simulated stack were set equal to those of Reference \cite{Thiesbrummel2024}. In the stack used in reference \cite{Thiesbrummel2024}, the PFN-Br layer ensures close interfacial contact and achieve exceptional crystallization \cite{Wang2024}. The BCP layer is a buffer between the electron transport layer and the metal electrode. Neither PFN-Br nor BCP layers are explicitly considered in our model as they do not influence the optical properties due to their narrow widths (less than 5 nm and 8 nm respectively). Our model considers [6, 6]-phenyl-C61-butyric acid methyl ester (PCBM) rather than C60 as PCBM was used by Giliberti {\it et al} in \cite{Magliano2025}. We took optical data from that paper and another by  Giliberti {\it et al} \cite{Giliberti2026} in the absence of refractive index data for the materials used by Thiesbrummel {\it et al} \cite{Thiesbrummel2024}. These papers considered device stacks with similar materials, including a perovskite with a nearly identical perovskite bandgap, 1.61 vs 1.63 eV (which is key, since the bandgap is the main factor determining the refractive indices).

$A$, the number of photons per area absorbed in the perovskite per second calculated by the TMM, is 1.16$\times$10$^{21}$m$^{-2}$s$^{-1}$, too low to be compatible with measured $J_{sc}$ values via $J_{sc} = q\times IQE \times A$, where $q$ is the absolute electronic charge and $IQE$ the internal quantum efficiency. This value of $A$ provides an upper limit for $J_{sc}$ of about 19 mA/cm$^2$ when $IQE$=1 that is substantially lower than the experimentally measured values of up to 23 mA/cm$^2$. Possible explanations for this result are the following. (i) In the visible spectrum for a device stack with ideal planar surfaces and interfaces, the TMM predicts a large fraction of photons either being reflected at layer interfaces or absorbed in the layers between the illuminated surface and the perovskite layer. These effects can both be reduced by surface roughness, given that perovskite grain sizes are of the order of visible wavelengths \cite{Dey2024,Jin2022}. Berry {\it et al} \cite{Berry2022} have demonstrated that patterning at the wavelength scale is an efficient way of increasing absorption, being able to $\approx$ 15 \% increase in external quantum efficiency over ideal planar surfaces.  (ii) Uncertainties may also come from surface treatments and decoherence due to encapsulating layers, not taken into account in the optical model. (iii) The refractive indices and layer thicknesses of the devices in reference \cite{Thiesbrummel2024} may differ from the values used here.

We therefore apply a constant multiplier, $G_{adj}$ = 1.184, to the calculated profiles to make the simulated results compatible with the measured $J_{sc}$ values. The value of this multiplier was determined by a BPE analysis of the pristine device (Age-0-min), where the multiplier was included as one of the BPE variables alongside the other inputs described below. $G_{adj}$ was kept constant in the analyses described below, since the charge carrier generation profile did not change as the device aged, according to the absorption spectroscopy measurements in \cite{Thiesbrummel2024} Supporting Information.

\begin{figure}
    \centering
    \includegraphics[width=1\linewidth]{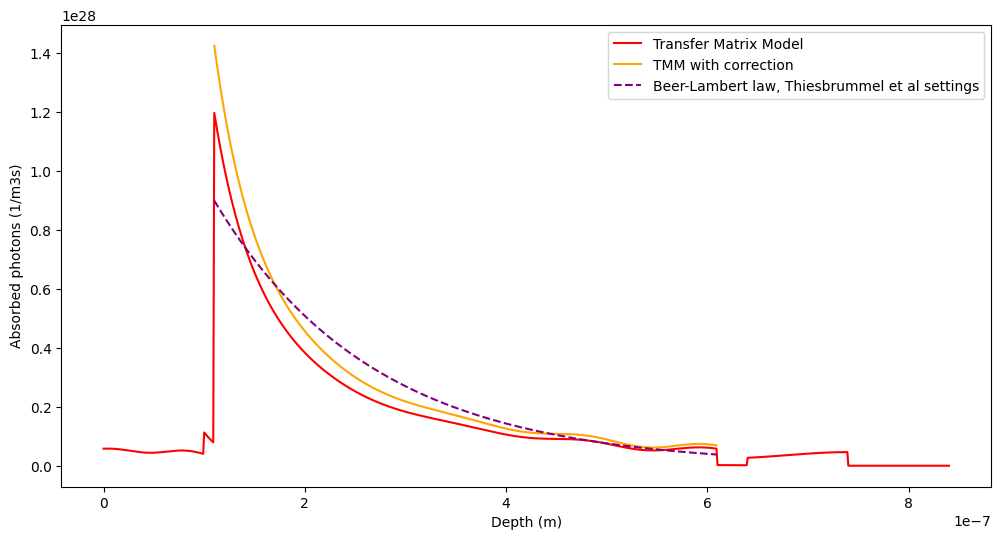}
    \caption{Photon absorption profiles. The red line shows the uncorrected TMM results for the full device stack. The orange line is the corrected TMM profile for the perovskite layer. This is compared to the default Beer-Lambert profile of IonMonger with inputs as used by Thiesbrummel et al (incident photon flux $F_{ph}$=1.42$\times$10$^{21}$ m$^{-2}$s$^{-1}$, perovskite absorption coefficient $\alpha$=6.34m$^{-1}$). }
    \label{fig:absprof}
\end{figure}

Figure \ref{fig:absprof} compares uncorrected and corrected TMM absorption profiles with a Beer-Lambert profile, the latter using values of the incident photon flux ($F_{ph}$) and perovskite absorption coefficient ($\alpha$) in \cite{Thiesbrummel2024}. The corrected TMM profile is higher at the HTL interface, but decreases more sharply than the Beer-Lambert profile and levels off, hinting at interference oscillations deeper in the perovskite, in contrast to the exponential decay of the Beer-Lambert profile.

\pagebreak
\subsection{Machine Learning Methods}
\label{sec:SIML}

We used the Metropolis-Hastings (MH) Markov Chain Monte Carlo (MCMC) method \cite{Gelman2013} in Bayesian Parameter Estimation (BPE)  to establish the distribution of IonMonger input parameters  consistent with the experimental JV-scan results. BPE applies Bayes' Rule to inversely derive a posterior distribution $p(\theta|\bf y_m)$ over a set of parameters $\theta$, associated with a measured outcome $\bf y_m$, given a prior distribution over $\theta$ and the likelihood of obtaining the measured outcome $\bf y_m$ from $\theta$, $p({\bf y_m}|\theta$). In our case the measured outcome is the set of experimental JV-scan results, specifically $J_{sc}$ (short-circuit current), $V_{ocr}$ (open-circuit voltage, reverse scan), $V_{ocf}$ (open-circuit voltage, forward scan), $\eta_r$ (power-conversion efficiency, PCE, reverse scan), and $\eta_f$ (PCE, forward scan), at 9 scan rates \cite{Thiesbrummel2024}, so $\bf y_m$ is a vector length 45. 

The vector $\theta$ is a set of inputs to the IonMonger simulation model, representing materials and device parameters. These inputs are a subset of the full set of IonMonger inputs, with those not included in the BPE being held constant. The relationship between the actual inputs (i.e.numerical values of the inputs) to IonMonger, $\hat{\theta_i}$ is specified flexibly with a particular element $\theta_i$ being either equal to $\hat{\theta_i}$ or log10($\hat{\theta_i}$), the latter option relevant to the cases where model inputs may vary over several orders of magnitude.

The likelihood of obtaining a measured result $\bf y_m$ is determined from running IonMonger for a given set of parameters $\theta$ to produce a prediction ${\bf y}$. The likelihood that $\theta$ produces the result $\bf y_m$ is modelled as a multivariate Normal distribution of the scaled prediction: $\tilde{\bf y} = {\bf y}/\bf y_m$ (element-wise) with a diagonal covariance matrix, with all diagonal elements set equal to a value $\rho$. This gives a formula for the log-likelihood (ignoring the constant normalisation term)
$L(\tilde{\bf y}) = - WMSE/2\rho$, where the  weighted mean-square-error is given by 
$WMSE = (1/N)\sum{{\bf w}(\tilde{\bf y}-1)^2}$, where $N$ is the number of elements in $\tilde{\bf y}$ and ${\bf w}$ is an array of weights, set to 2 for the $J_{sc}$ and $V_{ocf}$ terms, 1 otherwise, in an ad-hoc adjustment giving greater weight to quantities that are particularly sensitive to the model inputs. 

We assume independent uniform priors within specified boundaries for each input. The variable space for the MCMC trajectories apply a linear scaling, with boundaries 0 and 1 for each scaled variable. The MCMC trajectories are defined in this scaled parameter space. Each chain starts from a random point in the parameter space (as specified by the uniform priors) and proceeds in
steps with normal distribution defined by $\sigma_{step}$. For each new proposed location in parameter space IonMonger is run to get JV-scan result to be compared to the target, as detailed above. Proposed steps are accepted or rejected according to the Metropolis-Hastings algorithm \cite[Chapter 11.2]{Gelman2013}. The presence of both spatial and temporal stiffness in the model equations makes the task of finding solutions to realistic models of PSCs very challenging. The design of IonMonger \cite{Courtier2018} means that it is generally able to overcome these difficulties, but in the small fraction of cases where IonMonger fails to find a solution for a particular combination of inputs, this is treated as a rejected step in the MH procedure.

For the BPE hyper-parameters we used a value of $\sigma_{step}$ = 0.1.
The parameter $\rho$, the variance in the log-likelihood denominator, was adjusted to give average acceptance rates of the proposals in the MCMC chains in the range 0.2-0.31, close to the optimal jumping rule in high dimensions of 0.23 \cite[Chapter 12.2]{Gelman2013}. The lengths of chains were always above 200, where the first half of each chain was discarded when calculating posterior distributions and the numbers of chains always greater than 20. See the description of individual results for the exact values used for each simulation.

The core BPE analysis focused on 6 input parameters: $N_0$, $D_I$, $v_{nE}$, $v_{pE}$, $v_{nH}$, and $v_{pH}$. Table \ref{tab:BPEparams} shows the parameters used in the BPE for the different device ages and the average acceptance rate of jumps across the chains. 

Table \ref{tab:BPEres} gives the summary results of the BPE analyses for all device ages. The Gelman-Rubin convergence criterion ($\hat{R}$) \cite[Chapter 11]{Gelman2013} for the individual inputs are in the range 1.0-1.4, close to the ideal value of 1.0 (see Table \ref{tab:BPEres}). The Effective Sample Size \cite[Chapter 11]{Gelman2013} is greater than 200 for all variables in all four BPE analyses.

\begin{table}
\begin{tabular}{lrrrrr}
\hline 
{\bf Age} & {\bf $\sigma_{step}$} & {\bf $\rho$} & {\bf chain length} & {\bf number of chains} & {\bf acceptance rate} \\
\hline 
0 min & 0.1 & 0.0001 & 400 & 40 & 0.20 \\
90 min & 0.1 & 0.001 & 400 & 40 & 0.31 \\
280 min & 0.1 & 0.001 & 400 & 40 & 0.27 \\
480 min & 0.1 & 0.001 & 400 & 40 & 0.20 \\
\hline
\end{tabular}
\caption{\label{tab:BPEparams} BPE parameters and acceptance rates.}
\end{table}

\begin{table}[p]
\begin{tabular}{lrrrrr}
\hline 
{\bf Input} & {\bf BPE mean} & {\bf BPE std.dev.} & {\bf ESS} & {\bf std.err.} & {\bf Gelman-Rubin}\\ \hline
\multicolumn{6}{l}{ {\bf Age-0-min}} \\ \hline
$N_0$ & 21.6 & 0.77 & 347 & 0.04 & 1.38 \\
$D_I$ & -12.6 & 1.04 & 257 & 0.07 & 1.21 \\
$v_{nE}$ & 1.03 &  & 175 &  & 1.19 \\
$v_{pE}$ & 1.24 &  & 252 &  & 1.26 \\
$v_{nH}$ & 0.48 & 0.58 & 316 & 0.03 & 1.20 \\
$v_{pH}$ & 0.74 & 1.08 & 223 & 0.07 & 1.28 \\ \hline

\multicolumn{6}{l}{ {\bf Age 90 min}} \\ \hline
$N_0$& 23.2 & 1.21 & 206 & 0.09 & 1.19 \\
$D_I$& -13.7 & 1.07 & 228 & 0.07 & 1.13 \\
$v_{nE}$& 0.95 &  & 215 &  & 1.10 \\
$v_{pE}$& 1.15 &  & 252 &  & 1.18 \\
$v_{nH}$& 0.71 & 0.78 & 256 & 0.05 & 1.18 \\
$v_{pH}$& 0.47 & 1.08 & 269 & 0.07 & 1.22 \\ \hline

\multicolumn{6}{l}{ {\bf Age 280 min}} \\ \hline
$N_0$& 23.3 & 1.11 & 228 & 0.07 & 1.27 \\
$D_I$& -13.6 & 0.98 & 245 & 0.06 & 1.20 \\
$v_{nE}$& 1.10 &  & 236 &  & 1.15 \\
$v_{pE}$& 0.91 &  & 237 &  & 1.13 \\
$v_{nH}$& 0.91 & 0.72 & 267 & 0.04 & 1.15 \\
$v_{pH}$& 0.45 & 1.07 & 263 & 0.07 & 1.34 \\ \hline

\multicolumn{6}{l}{ {\bf Age-480-min}} \\ \hline
$N_0$& 23.9 & 1.07 & 272 & 0.07 & 1.37 \\
$D_I$& -14.2 & 0.97 & 296 & 0.06 & 1.25 \\
$v_{nE}$& 0.89 &  & 224 &  & 1.31 \\
$v_{pE}$& 1.06 &  & 200 &  & 1.37 \\
$v_{nH}$& 1.21 & 0.63 & 307 & 0.04 & 1.27 \\
$v_{pH}$& 0.37 & 0.89 & 256 & 0.06 & 1.36 \\ \hline

\hline
\end{tabular}
\caption{\label{tab:BPEres} BPE results. Standard deviations and effective sample sizes, ESS, are not shown for $v_{nE}$ and $v_{pE}$ since the distributions do not possess a clearly defined peak and the calculated standard deviations are indicative more of the prior ranges than of the true posterior distributions.}
\end{table}

\begin{figure}[p]
    \centering
    \includegraphics[width=1\linewidth]{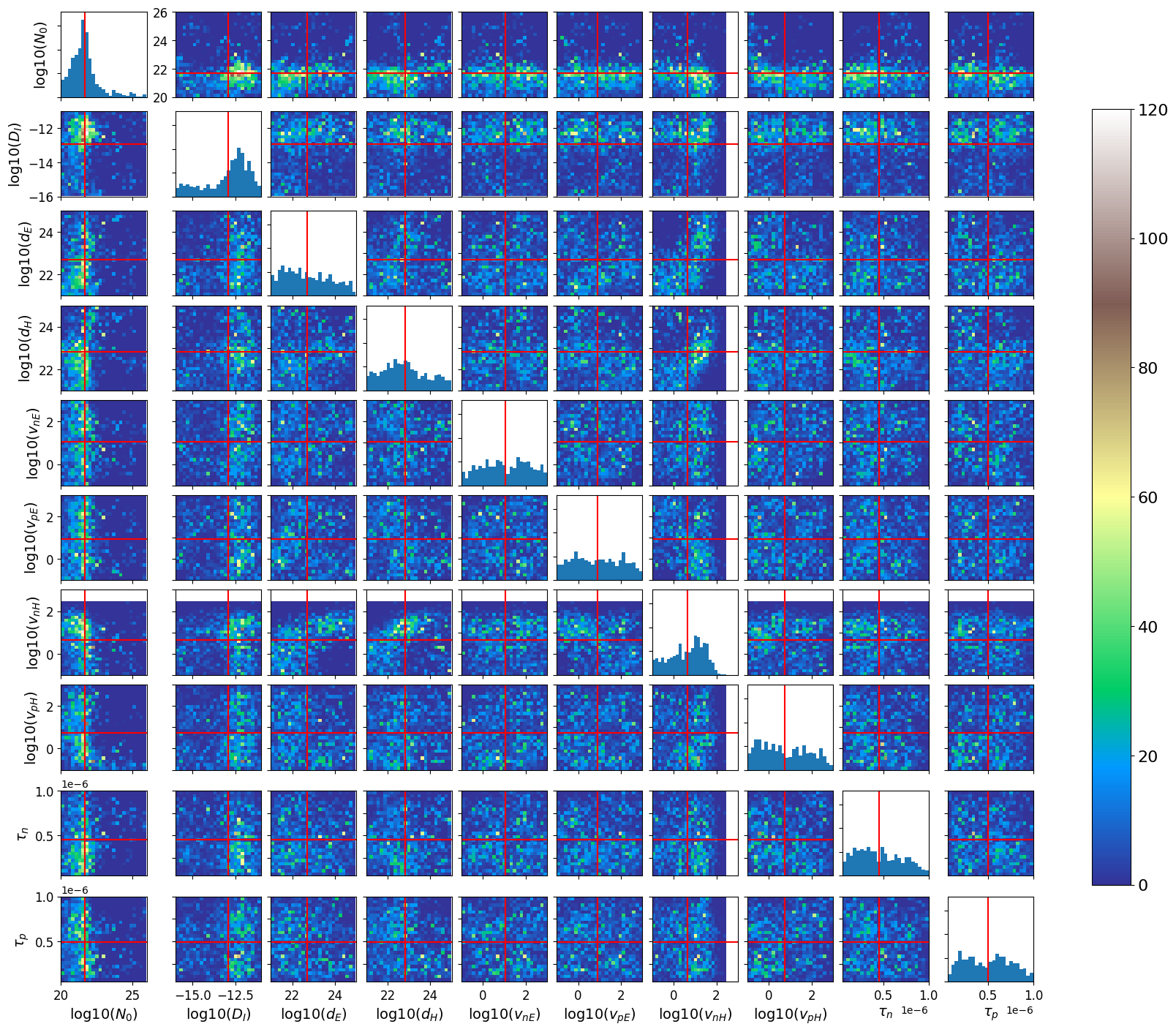}
    \caption{Posterior distributions at age 0 min from BPE for $N_0$, $D_I$, $d_E$, $d_H$ , $v_{nH}$, $v_{pH}$, $v_{nH}$, $v_{pH}$,
     $\tau_n$, $\tau_p$ . Single parameter distributions are shown along the diagonal and joint two-parameter distribution as heat maps off-diagonal. The red lines show the means of the single parameter distributions.  For $\tau_n$ and $\tau_p$  the BPE uses the actual values, for the other inputs it uses log10 of actual inputs. BPE settings are $\sigma_{step}$=0.1, $\rho$=0.0001, 49 chains of 200 steps, discarding first half of each chain.}
    \label{fig:age0_10vars}
\end{figure}

\begin{figure}[p]
    \centering
    \includegraphics[width=1\linewidth]{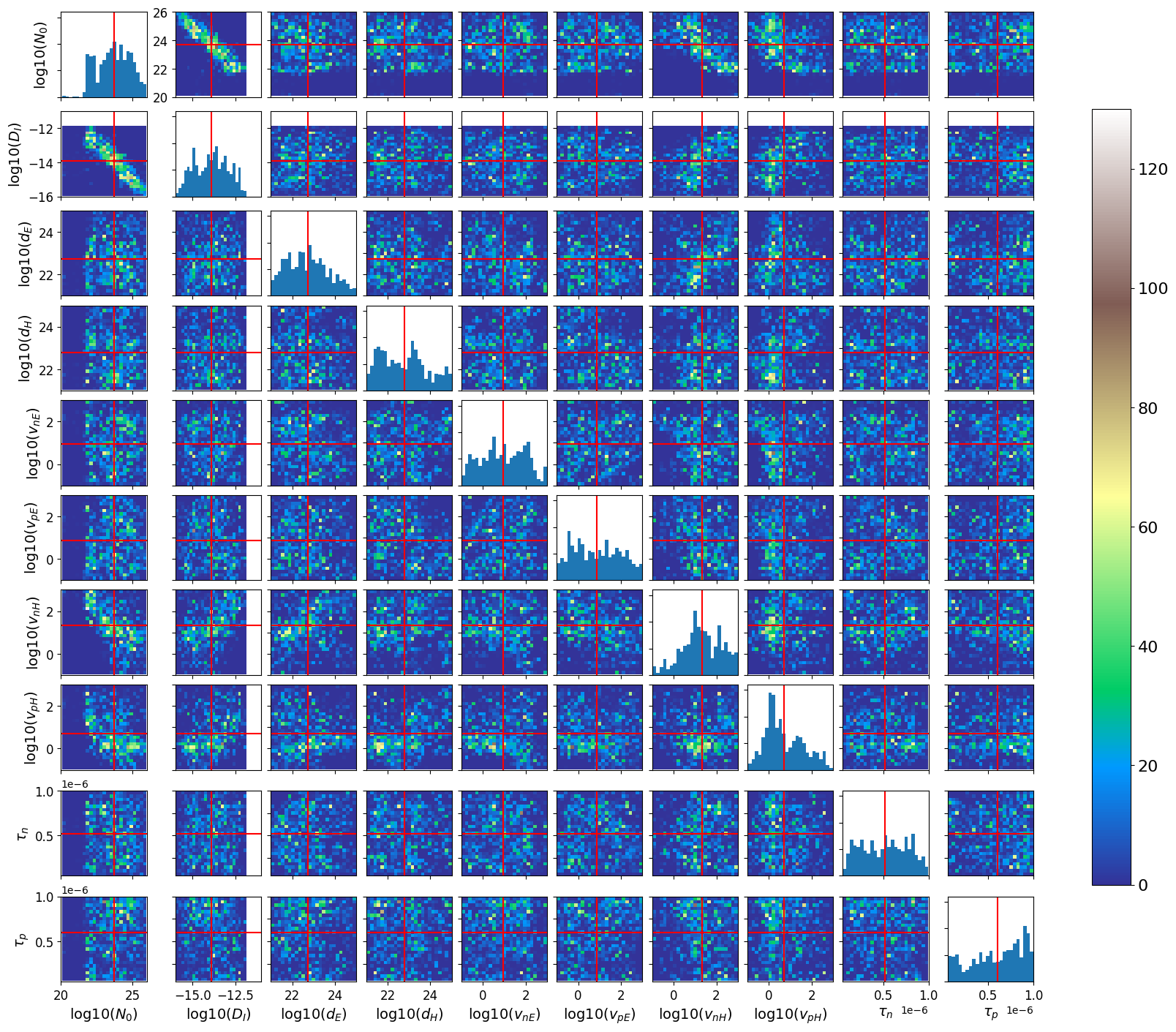}
    \caption{Posterior distributions from BPE at age 480 min. See caption to Figure \ref{fig:age0_10vars} for description. BPE settings: $\sigma_{step}$=0.1, $\rho$=0.001, 50 chains of 200 steps, discarding first half of each chain.}
    \label{fig:age480_10vars}
\end{figure}

Figures \ref{fig:age0_10vars} and \ref{fig:age480_10vars} show single-parameter and joint pair-wise posterior distributions from the initial BPE analysis with 10 inputs: mobile ion density $N_0$, diffusion coefficient $D_I$; CTL doping densities, $d_E$, $d_H$; interface recombination velocities at the ETL interface $v_{nE}$, $v_{pE}$, and at the HTL interface $v_{nH}$, $v_{pH}$; and carrier lifetimes $\tau_n$ and $\tau_p$. These figures show that the values of the CTL doping densities, $d_E$ and $d_H$, and the carrier lifetimes, $\tau_n$ and $\tau_p$ do not make a significant difference to the JV-scan results, justifying fixing these parameters at the values in Table 1 of the main paper. 

\newpage

\subsection{Photoluminescence}
\label{sec:SIPL}

We predict the Quasi-Fermi-Level Splitting (QFLS, $\Delta\mu$), which is the difference between quasi Fermi levels for electrons and holes, which in turn are related to the electron and hole concentrations, respectively, using IonMonger with open-circuit boundary conditions \cite[Section 3.4]{Courtier2018}. $\Delta\mu$ is related to the PL quantum yield, PLQY, as follows \cite{Stolterfoht2019}
\begin{equation}
\Delta\mu = k_B T \ln \left( \frac{J_{rad}}{J_{0,rad}}\right) 
= k_B T \ln \left( \text{PLQY}\frac{J_G}{J_{0,rad}}\right) 
\label{eq:s3eq1}
\end{equation}
Here $J_G$ is the total recombination current accounting for radiative and non-radiative losses and $J_{0,rad}$ is the radiative thermal equilibrium recombination current density in the dark.
 
The results from the PL measurements of Ref. \cite{Thiesbrummel2024} on a device stack similar to the device used for the JV-scans are presented in terms of QFLS values. The result for the pristine device shows a value of QFLS very close to the open-circuit voltage, $V_{oc}$, but as the device ages, a growing gap between QFLS and $V_{oc}$ opens up, which can be caused by the presence of mobile ions or a range of interface-related effects \cite{Warby2023,Kober-Czerny2025}. These PL experiments used a monochrome incident light with wavelength 520 nm with the intensity adjusted to a 1 sun equivalent intensity. Our combined optical and transport model enables us to model QFLS with monochrome illumination. The intensity was adjusted to produce a theoretical $J_{sc}$ of 22 mA/cm$^2$, in a similar way to how the intensity was adjusted in the experimental procedure in Ref. \cite{Thiesbrummel2024}. As stated in Ref \cite{Thiesbrummel2024} a $J_{sc}$ of 22 mA/cm$^2$ is equivalent to 1.375$\times$10$^{21}$ photons m$^{-2}$s$^{-1}$, assuming perfect carrier extraction and {\it that this is the number of photons absorbed in the perovskite layer}. However, our optical model predicts that at 520 nm there is about 25\% reflection, so if 1.375$\times$10$^{21}$ photons m$^{-2}$s$^{-1}$ are to be absorbed in the perovskite (to achieve a $J_{sc}$ of 22 mA/cm$^2$), the incident flux is set to 1.835$\times$10$^{21}$ photons m$^{-2}$s$^{-1}$. The results presented below and in the main paper use this assumption for the incident flux. To test the sensitivity to this assumption, we also did simulations with an incident flux of 1.375$\times$10$^{21}$ photons m$^{-2}$s$^{-1}$. These simulations produced nearly identical results, but with $qV_{oc}$ and $\Delta\mu$ shifted downwards by a constant amount of about 0.01 eV. 

\begin{figure}
    \centering
    \begin{subfigure}[b]{0.49\textwidth}
        \centering
        \includegraphics[width=\textwidth]{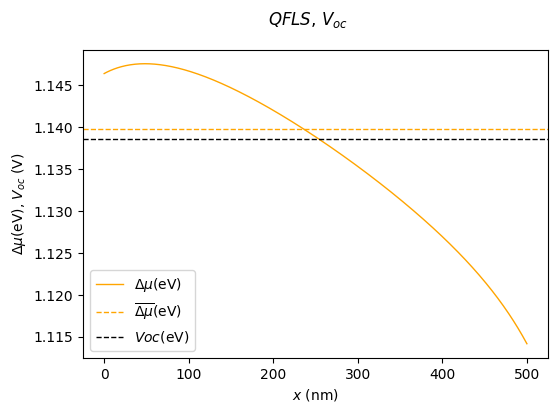} 
        \caption{}
    \end{subfigure}
    \hfill
    \begin{subfigure}[b]{0.49\textwidth}
        \centering
        \includegraphics[width=\textwidth]{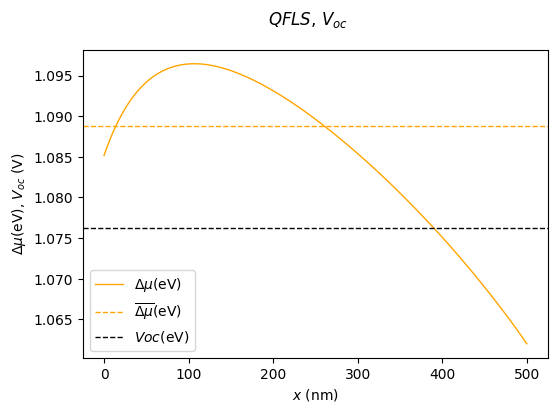}
        \caption{} 
    \end{subfigure}
    \caption{Quasi-Fermi-Level Splitting (QFLS, $\Delta\mu$) across the perovskite layers from open-circuit runs of IonMonger with inputs as the BPE means from the analyses for Age-0-min (Panel (a)) and Age-480-min (Panel (b)). The horizontal lines show the means of QFLS across the perovskite layer and the value of $V_{oc}$ for the device.}
    \label{fig:Voc_QFLS_profile}
\end{figure}

Figure \ref{fig:Voc_QFLS_profile} shows the profiles of QFLS across the perovskite layer ($\Delta\mu(x)$) for open-circuit IonMonger simulations with inputs set to the BPE means from the Age-0-min and Age-480-min analyses (see Table 1 of the main paper). The profiles are not uniform, both due to the non-uniform carrier-generation profile (heavily weighted towards the HTL side) and movement of carriers within the device. The figure also shows the weighted means of the QFLS across the perovskite ($\overline{\Delta\mu}$) and $V_{oc}$ as horizontal lines. $\overline{\Delta\mu(x)}$ is calculated as the mean of the $\Delta\mu(x)$ profile weighted by the profile of radiative recombination by depth, which is proportional to $\exp(\Delta\mu(x)/k_BT)$. Thus, from eq (\ref{eq:s3eq1}),
\begin{equation}
\overline{\Delta\mu} = \frac{\int_0^b{\Delta\mu(x)\exp(\Delta\mu(x)/k_BT)dx}}{\int_0^b{\exp(\Delta\mu(x)/k_BT)dx}}
\end{equation}
where the integrals are over the width of the perovskite layer, from depth 0 to the thickness $b$, $k_B$ is Boltzmann's constant and $T$ is the temperature.
The QFLS means and $V_{oc}$ for all four ages in the BPE analyses are shown in panel (b) of Figure 6 in the main paper, demonstrating the growing difference between QFLS and $V_{oc}$ as the device ages. 


The results presented here are in qualitative agreement with Thiesbrummel {\it et al} \cite{Thiesbrummel2024} in that the QFLS-$V_{oc}$ mismatch grows as the device ages. However, there are quantitative differences. In the simulations both QFLS and $V_{oc}$ decrease with age and the difference at age 480 min is 0.012 eV. In contrast, Ref. \cite{Thiesbrummel2024} reports QFLS growing with age with the QFLS-$V_{oc}$ gap at an age near 480 min of about 0.1 eV, i.e. an order of magnitude greater than the gap in the simulation results. 


To understand how a wider range of interface recombination velocities influences QFLS, $V_{oc}$, and the difference between them, we did an additional set of open-circuit IonMonger runs, varying $v_{nE}$, $v_{pE}$, $v_{nH}$, $v_{pH}$. Each of these parameters is set to three values: 10$^{-3}$ ms$^{-1}$, 10 ms$^{-1}$ (close to the Age-480-min BPE means), and 10$^3$ ms$^{-1}$. Simulations were run for all combinations of these values. These runs were made with the $N_0$ and $D_I$ values from the Age-480-min BPE (see Table \ref{tab:BPEres}), so the results would be representative of the state of the degraded device.

\begin{figure}[p]
    \centering
    \begin{subfigure}[b]{0.49\textwidth}
        \centering
        \includegraphics[width=\textwidth]{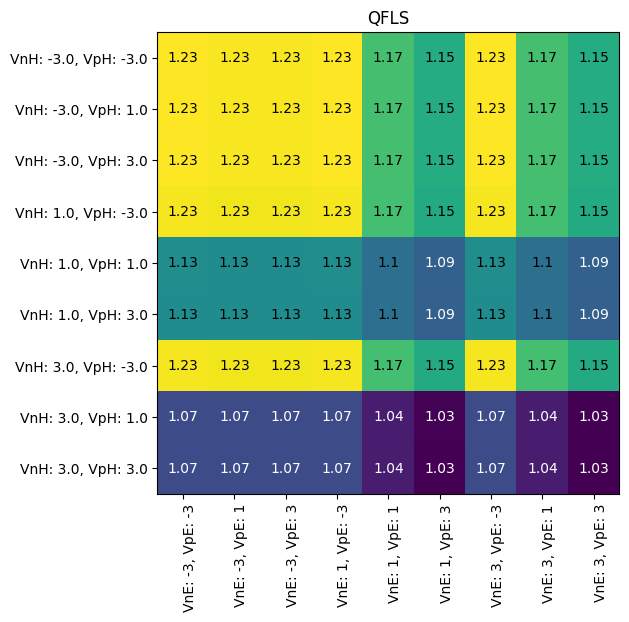} 
        \caption{}
    \end{subfigure}
    \hfill
    \begin{subfigure}[b]{0.49\textwidth}
        \centering
        \includegraphics[width=\textwidth]{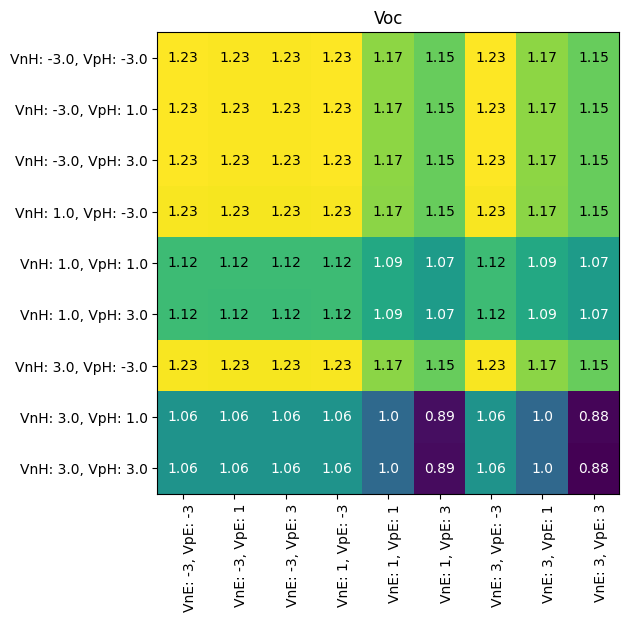} 
        \caption{}
    \end{subfigure}
    \hfill
    \begin{subfigure}[b]{0.49\textwidth}
        \centering
        \includegraphics[width=\textwidth]{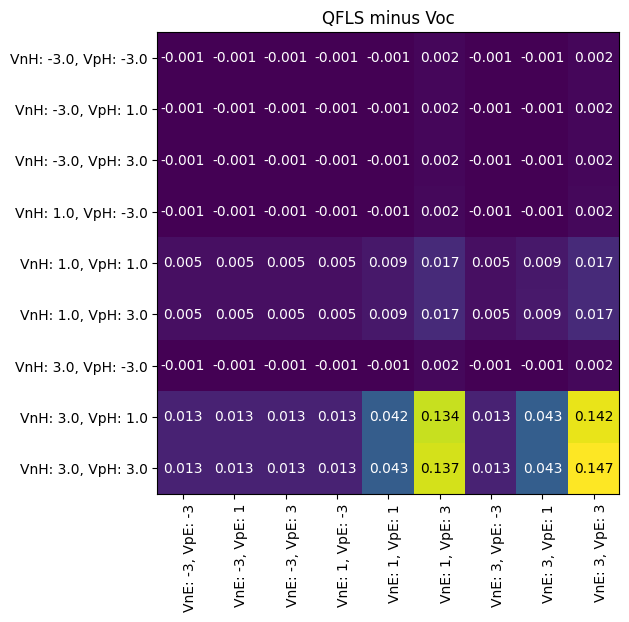}
        \caption{} 
    \end{subfigure}
    \caption{QFLS, $qV_{oc}$ and their difference, in eV, from open-circuit IonMonger runs for combinations of interface recombination velocity values. The values shown in the axis labels are the model inputs, i.e. log10 of the physical values. Panel colors aid visualisation of the value shown by the numbers in the panels}
    \label{fig:QFLS_vloop}
\end{figure}

The results are shown in Figure \ref{fig:QFLS_vloop} from which the following conclusions may  be drawn
\begin{itemize}
\item QFLS can be higher than that of the pristine (Age-0-min) device, 1.14 eV, at low values of $v_{nH}$ and/or $v_{pH}$ (top rows of panel (a)), but in all these cases the difference between QFLS and the values of $qV_{oc}$ shown in panel (b) is tiny, $\approx$ 0.001 eV (top parts of panel (c)).
\item A value of QFLS-$qV_{oc}$ exceeding 0.1 eV is obtained with high values for all interface recombination velocities (bottom-right cells of panel (c)), but in these cases both QFLS and $qV_{oc}$ are substantially below the value of 1.23 eV for the pristine device (bottom-right cells of panels (a) and (b)).
\end{itemize}
None of these simulations produce the simultaneous high QFLS and large gap between QFLS and $qV_{oc}$ presented in Refs. \cite{Thiesbrummel2024,Warby2023}. Those results therefore cannot be explained by changes in $N_0$, $D_I$ or the recombination velocities. There are differences between the devices used for JV-scans and PL, which may hinder direct comparisons \cite{Aalbers2025} and they may have been aged under different conditions, which could mean that the parameters derived from the JV-scans may not accurately reflect the condition of the PL devices. Other effects, such as changes to interlayers or adverse changes to the CTL-perovskite band alignments, could also influence the PL results, while having less impact on JV-scans \cite{Aalbers2025, Ceratti2026}.  Another potential complication is highlighted by Kalasariya \textit{et al} \cite{Kalasariya2026} who showed that for mixed-halide perovskites phase segregation, where halide atoms segregate into regions dominated by one or other of the halide constituents, can occur on similar time scales to those analysed in this paper, i.e. $<$1000 minutes, and that this can have a marked influence on measured PL spectra. 

\newpage
\subsection{Impact of model input changes on JV scan results}
\label{sec:SIIM}

\begin{figure}[p]
\includegraphics[width=1.0\linewidth]{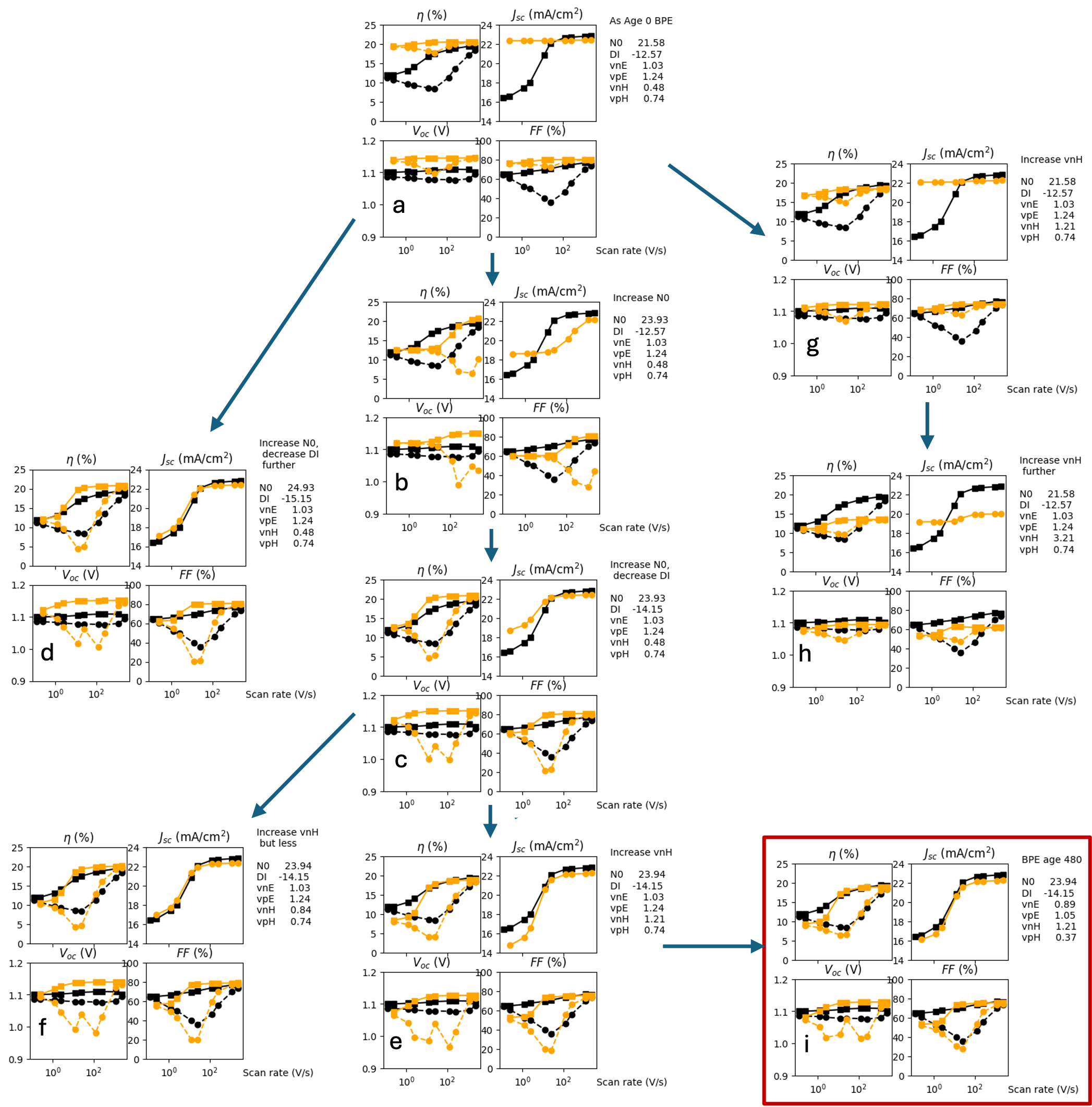} 
\caption{IonMonger results for different input combinations, evolving from the Age 0 (top panel) to the Age-480-min BPE results (bottom-right panel), shown as orange lines and symbols, compared to experimental results, black lines and symbols, from the Age-480-min device (the same in all panels). Squares show reverse scans and discs show forward scans.  The values for the IonMonger inputs generating the model results in each panel are shown alongside the charts of PCE ($\eta$), $J_{sc}$, $V_{oc}$ and $FF$ against scan rate. The input values shown are all log10 of the model inputs; thus, for example, N0 of 21.58 corresponds to $N_0$ = 3.8$\times$10$^{21}$ m$^{-3}$, and vnH of 0.48 to $v_{nH}$ = 3.02 m/s.}
\label{fig:age0_to_age480}
\end{figure}

Figure \ref{fig:age0_to_age480} demonstrates the impact of the changes to the different input parameters to IonMonger, when moving from the BPE means found for the Age-0-min device to those found for the Age-480-min device. The experimental results in all panels are those of the Age-480-min device. 

The logical sequence of model runs shown in the different panels is

\begin{itemize}
\item Panel a: The model results are generated with the inputs found for the BPE for the Age-0-min device. These model results are clearly very different from the experimental results for the Age-480-min device.
\item Panel b: For this model run N0 has been increased to the value from the Age-480-min BPE, but other inputs have been kept as the Age-0-min inputs. This increase produces a substantial drop in $J_{sc}$ at low scan rates and increased hysteresis in $\eta$ and $FF$. However, the scan rate at the steepest gradient in $J_{sc}$ is too high.
\item Panel c: Decreasing DI, as well as increasing N0, (both set to the Age-480-min BPE values) corrects the scan-rate misalignment seen in Panel b. However, the $J_{sc}$ at low scan rates is higher than seen in the experiments.
\item Panel d: One possibility for fixing this problem (too high $J_{sc}$ at low scan-rate) is to increase N0 and simultaneously decrease DI (to ensure correct positioning along the scan-rate axis) even further, beyond the Age-480-min BPE values. This produces quite a reasonable fit to the experimental results, certainly for $J_{sc}$. However, the hysteresis seen in the $\eta$ and $FF$ subpanels is more severe than seen in the experiments.
\item Panel e: An alternative to the further changes in N0 and DI is to explore the impact of changes in the interface recombination velocities, in particular vnH. The results in this panel are from a run with the same N0 and DI value as Panel c, but increasing vnH to the value found in the Age-480-min BPE. This increase in vnH causes an overshoot in the low-scan-rate drop in $J_{sc}$, predicting too low values. But this panel clearly demonstrates the sensitivity of the results to the value of vnH.
\item Panel f: One possible fix for the low-scan-rate $J_{sc}$ undershoot in Panel e would be to increase vnH by a smaller amount, here chosen as a value half-way between the Age-0-min and Age-480-min BPE results. This does produce a good fit for $J_{sc}$, but the hysteresis in $\eta$ and $FF$ is predicted to be stronger than observed experimentally, as was also the case for the model results in Panel d.
\item Panel g: It is instructive to explore what happens when changing only vnH while keeping N0 and DI at the Age-0-min level. This panel shows that changing from the Age-0-min to the Age-480-min BPE values for vnH has practically no effect on the JV results, at this N0 value of 3.8$\times$10$^{21}$ m$^{-3}$ .
\item Panel h: Increasing vnH further, from the value of 1.21 in Panel g to 3.21, increasing $v_{nH}$ by two orders of magnitude, does change the characteristics, but not in agreement with the experiments. $J_{sc}$, $\eta$ and $FF$ for a reverse scan drop at all scan rates, not just low ones, as was also observed by Thiesbrummel {\it et al} \cite{Thiesbrummel2024} in their simulations. This observation emphasises that the strong effects of changing vnH, observed in Panels e and f, are only seen in the presence of high concentrations of mobile ions.
\item Panel i: Whilst changing just one of the recombination velocity inputs, i.e. vnH, can produce results quite close to experiments, as shown in Panel f, the BPE is able to achieve a better overall fit when also varying the other three recombination velocity inputs, vpH, vnE and vpE, but all by smaller amounts than the change in vnH. This panels shows the run with the values for all six inputs set to the means of the BPE for the Age-480-min experiments. The agreement for $J_{sc}$ is similar to what is seen in Panels d and f, but this run reproduces the amount of hysteresis seen in the $\eta$ and $FF$ subpanels better than the results in Panels d and f.  
\end{itemize}

\end{document}